\documentclass{ws-ijmpe}
\usepackage[super,compress]{cite}
\usepackage{epsfig}
\usepackage{multirow}
\begin{document}



\title{Multiplicity spectra in $e^{+}e^{-}$ and $\overline{p}p$ collisions in terms of 
      Tsallis and Weibull distributions.}
\author{S. Sharma, M. Kaur, S. Thakur}
\address{Department of Physics, Panjab University, Chandigarh\\
India\\
manjit@pu.ac.in\footnote{corresponding author}}
\maketitle
\begin{abstract}
Charged hadron production in the $e^{+}e^{-}$ annihilations at 91 to 206 GeV in full phase space and in $\overline{p}p$ collisions at 200 to 900~GeV collision energies are studied using non-extensive Tsallis and stochastic Weibull probability distributions.~The Tsallis distribution shows better description of the data than the Weibull distribution. The 2-jet modification of the statistical distribution is applied to describe $e^{+}e^{-}$ data.~The main features of these distributions can be described by a two-component model with soft, collective interactions at low transverse energy and hard, constituent interactions dominating at high transverse energy.~This modification is found to give much better description than a full-sample fit, and again Tsallis function is found to better describe the data than the Weibull one pointing at the non-extensive character of the multiparticle production process.

\end{abstract}  
\ccode{PACS numbers: 5.90.+m, 12.40.Fe, 13.66.Bc}

\section{Introduction}
Charged hadron production in high energy particle interactions can be understood in terms of several theoretical and phenomenological models derived from statistics, empirical relationships, phenomenological concepts or pure theoretical cosiderations [\refcite{Kanki}-\refcite{Gros}].~To understand the particle production mechanism, some models have used ensemble theory from statistical mechanics to extract information from dynamical fluctuations as a key source of inputs to study the multiplicity patterns.~Several distributions based on the statistical analyses have been derived for the understanding of particle production and the multiplicity distributions.~However, it was realized that data on many single particle distributions deviate from the distributions expected from the statistical models, based on standard statistical mechanics of Boltzman-Gibbs which treats the entropy as an extensive property. This initiated the idea of including modifications to include possible intrinsic, non-statistical fluctuations. These were identified as the source of the deviations.~Such fluctuations are important as possible signals of phase transition(s) taking place in an hadronizing system.

The Tsallis formalism was introduced three decades ago [\refcite{TS1}].~It is a non-extensive generalization of the statistical mechanics and has been very successful in describing very different physical systems in terms of statistical approach, including multiparticle production processes at lower energies.~Specially designed to include self-similar systems and systems with long range interactions, it is a reasonable choice to study hadronisation process.~The collisions at collider experiments are expected to fall in this category.~Thus the Tsallis statistical approach [\refcite{TS2}-\refcite{JC}], was successfully used for multiparticle production processes.~The new approach in the Tsallis $q$-statistics which includes entropy as a non-extensive property to describe the particle production has not been tested with the highest energy data from $e^{+}e^{-}$ collisions and $\overline{p}p$ collisions, though this has been used to explain heavy-ion and some $pp$ collision data, see references [\refcite{Zhang}-\refcite{Mar}].~Most of these studies have focussed on the $p_{T}$ spectra of the hadrons.

~The non-extensive part of entropy is quantified in terms of a parameter $q$, the entropic index in the Tsallis function and envisaged that it should have a value greater than unity.~The total Tsallis entropy of two sub-systems $a$ and $b$ is not equal to the sum of the entropies of the subsystems, but is given by;
 \begin{equation}
 S_q(a,b)= S_a + S_b + (1-q)S_{a}S_{b}\, \label{one}
 \end{equation}
 
In Tsallis $q$-statistics probability is calculated by using the partition function Z, as  
\begin{equation}
P_N = \frac{Z^{N}_q}{Z}
\end{equation}
where Z represents  the total partition function and $Z^{N}_q$ represents partition function at a particular multiplicity, of the grand canonical ensemble of gas consisting of $N$ particles.~$\bar{N}$, the average number of particles, is given by 
 \begin{equation}
 \bar{N}=Vn[1 + (q-1)\lambda(V n\lambda-1)- 2v_0n] 
 \end{equation} 
 $K$-parameter is related to $q$ and excluded volume $v_{0}$, by
 \begin{equation}
\frac{1}{K}=(q-1)\lambda^{2} - 2\frac{v_0}{V} 
\end{equation} 
where $V$ is the volume containing $N$ particles, $\lambda$ is related to the temperature through the parameter $\beta$ as;
\begin{equation}
\lambda(\beta,\mu)= -\frac{\beta}{n}\frac{\partial n}{\partial\beta}
\end{equation} 

Details of the Tsallis distribution and how to find the probability distribution can be obtained from [\refcite{TS2}].
In the hadron-hadron interactions, the dynamics of particle production is centered around the fragmentation process, partons are produced at the intermediate stage which quickly undergo fragmentation into hadrons.

~A description of the fragmentation process can also be given in terms of Weibull distribution [\refcite{Wei}] in which it has been shown that result of a single event fragmentation leading to a branching tree of cracks in the materials that show fractal behaviour [\refcite{Brown}] and can be described by a Weibull-like distribution.~This is related to the particle number distribution developed during the fragmentation, the so called multiplicity distribution in case of particle collisions.
 
~Weibull distribution has been studied to describe multiplicity distributions in $e^{+}e^{-}$ collisions up to 91 GeV [\refcite{Wei}] and also for $ep$ collisions.~Weibull parametrization of the multiplicity distribution has also been used to describe the multi-dimensional fluctuations and genuine multi-particle correlations in $e^{+}e^{-}\rightarrow Z^{0} \rightarrow hadrons$ [\refcite{Edward},\refcite{Nayak}].

~The Tsallis distribution [\refcite{TS1},\refcite{TS2}] and the Weibull distribution [\refcite{Wei}] are based on the concepts of statistical mechanics and stochasticity.~The simplicity of these statistically inspired models and the ease of application to data, is the beauty of the analyses.~Nowadays the statistical approach is a standard procedure used to model high energy multiparticle production processes [\refcite{Ga}].

In the present work, we focus on the multiplicity distributions, in full phase space in $e^{+}e^{-}$ annihilations up to the highest available center-of-mass energies and in $\overline{p}p$ collisions at energies ranging from 200 to 900 GeV in restricted rapidity windows.~The presence of a shoulder structure in the multiplicity distribution was observed in $e^{+}e^{-}$ at collision energy of $\sqrt{s}$=91 GeV [\refcite{KT2},\refcite{OPAL91}] and in $\overline{p}p$ data at 900 GeV [\refcite{UA51}] and at $\sqrt{s}$=1800 GeV [\refcite{Rimo}].~The data from LEP2 also confirmed the presence of the shoulder structure at higher then $Z^{0}$ energies [\refcite{OPAL161},\refcite{OPAL172}].~It is the higher energy data where the shoulder structure becomes more pronounced.

~Recently, we have studied [\refcite{MK}] the data in terms of the Tsallis distribution and the Weibull distribution at lower energies in full phase space and in different rapidity windows.~To study the high energy data and to take into account the shoulder structure, we implement a 2-jet modification to improve the comparison between the predicted and the experimental values.~The two-component approach has been suggested earlier in reference [\refcite{Gov}].~The study indicates that the Tsallis function is able to reproduce the experimental results far better than the Weibull one.~Further analysis is performed here to assert our conclusions.

In Section 2, an outline of probability distribution functions of Tsallis and Weibull distrbutions and their modified forms are given.~Details along with the references can be found in [\refcite{TS2}].~2-jet fractions denoted by $\alpha$ at various energies have been taken from the references [\refcite{OPAL161},\refcite{OPAL172},\refcite{Alex}-\refcite{L3}] obtained by OPAL and L3/ALEPH experiments.~The 2-jet data samples considered are only those for $e^{+}e^{-}$ collisions, and no 2-jet modification is made to study $\overline{p}p$ data as no 2-jet data-samples being available for these collisions.~Section~3 presents the analyses of experimental data and the results obtained by the two approaches.~Discussion and conclusion are given in Section~4.
\section{Tsallis and Weibull distributions and their 2-jet modification}
\subsection{Tsallis and its 2-jet modification}
Details and the method of calculating the partition function for $N$ particles, in the Grand Canonical Ensemble, Tsallis $q$-index and Tsallis $N$-particle probability distribution can be obtained from [\refcite{TS2}].~The probability is calculated by using the partition function $Z$ , as  
\begin{equation}
P_N = \frac{Z^{N}_q}{Z}
\end{equation}
where $Z$ represents  the total partition function and $Z^{N}_q$ represents partition function at a particular multiplicity $N$.

The probability distribution for the 2-jet modified distribution is calculated by adding a weighted superposition of multiplicity in 2-jet and in multi-jet events as follows; 
\small
\begin{equation}
P_{N}(\alpha:\bar{n_1},V_1,v_{01},q_1:\bar{n_2},V_2,v_{02},q_2)= \\
\alpha P_{N}^{\rm 2-jet} +\\
(1-\alpha)P_{N}^{\rm multi-jet}.
\end{equation}
\normalsize
where $\alpha$ is a weight factor which gives 2-jet fraction and is determined from a jet finding algorithm. 

\subsection{ Weibull and its 2-jet modification} 
The probability density function of a Weibull random variable is;
\begin{equation}
 P_N(N,\lambda,k) = \frac{k}{\lambda} (\frac{N}{\lambda})^{(k-1)} exp^{-(\frac{N}{\lambda})^{ k}}, \hspace{0.4cm} where N \geq 0.                                   
\end{equation} 
 
Where $\lambda >0$ is the scale parameter and  $k>0$ is the shape parameter.~These two parameters for the distribution are related to the mean of the function, as 
\begin{equation}
\bar{N} = \lambda \Gamma(1+1/k)
\end{equation}
The modified Weibull function can be obtained by the weighted superposition of the two 
Weibull distributions, as above, namely;  
\small
\begin{equation}
P_{N}(\alpha:N_1,\lambda_1,k_1;N_2,\lambda_2,k_2)=\\
\alpha P_{N}^{\rm 2-jet} + \\
(1-\alpha)P_{N}^{\rm multi-jet}. 
\end{equation}
\normalsize

\section{Results}
The experimental data on $e^{+}e^{-}$ annihilation at different collision energies by two experiments are analysed here.~The details of the data from the L3 [\refcite{L3}] and OPAL [\refcite{OPAL91},\refcite{OPAL161},\refcite{OPAL172},\refcite{Alex},\refcite{Acc}] experiments at different energies between $\sqrt{s}$ = 91 to 206 GeV in the full phase space are given in Table~1.~For comparison, we also analyse data from $\overline{p}p$ collisions at energies ranging from 200 to 900 GeV in restricted rapidity windows from the UA5 collaboration [\refcite{UA51},\refcite{UA52}] as given in Table~2.~The experimental distributions are fitted with the predictions from two probability functions, as described above.
\subsection{one-component Tsallis versus Weibull functions}
The probability distribution for Tsallis function is calculated from equation (2) and  Weibull function from eqns. (8) and (9) and applied to fit the $e^{+}e^{-}$ data in full phase space, as shown in Figures~1 and 2.~Parameters of the fits and $\chi^{2}/ndf$ values are given in Table~2 and the corresponding p values are given Table~3.
~We find that overall, Weibull distribution fails to reproduce the data, particularly in the high multiplicity regions, while Tsallis distribution shows good fit in full phase space.

~One observes that the  $\chi^{2}/ndf$ values are significantly lower for the Tsallis fits in comparison to the Weibull fits.~This is true for all energies.~A careful examination of the $p$ values from Table~3 shows that for the data from L3 experiment at all energies from 130 to 206 GeV, the Weibull fits  are statistically excluded with $CL < 0.1 \%$.~While for the Tsallis fits for the data only at 200.2 and 206.2 are statistically excluded with $CL < 0.1 \%$ and is good for all other energies with $CL > 0.1 \%$. 
~For the data from OPAL experiment, the Weibull fit is statistically excluded systematically for all energies between 91 to 189 GeV with $CL < 0.1 \%$ except being good for energy 172 GeV.~Again Tsallis fit is excluded only for 91 GeV data and remains a good fit for all energies from 133 to 189 GeV with $CL > 0.1 \%$.~A comparison of the $\chi^{2}/ndf$ and $p$ values in Table~3 shows that $\chi^{2}/ndf$ values for the Tsallis distributions are lower by several orders, confirming that Tsallis distribution fits the data far better than Weibull distribution.

~Figures~3 and 4 show the Tsallis and Weibull fits to the data on $\overline{p}p$ collisions from UA5 collaboration at energies ranging from 200 to 900 GeV in restricted rapidity windows as well as in full phase space.~The parameters of the fits are listed in Table~4.~Again the Tsallis distribution shows very good fits at all energies and all rapidities with statistically significant $p$ values $\chi^{2}/ndf$ values in comparison to the Weibull distribution, as shown in Table~5.~In each of the data samples of $e^{+}e^{-}$, for the Weibull distribution, $\lambda$ values increase with energy and in case of $\overline{p}p$ collisions these increase with energy as well as with the size of the central rapidity interval.~Similarly for Tsallis distribution, the $q$ value which measures the entropic index, of the Tsallis statistics is consistently higher than 1 in each of the above mentioned cases.~This confirms the property of the non-extensivity in the data as proposed by the Tsallis q-statistics. 
\subsection{Modified Tsallis versus modified Weibull distributions} 

As it is observed above, in the case of the Weibull fits, the description is good enough only for the data at some energies, while it gets quite high $\chi^{2}/ndf$ values for the rest.~Also, though the Tsallis fits are better than the Weibull ones, there is still a room for improvements.~Following the two-component approach suggested by Giovannini [\refcite{Gov}] we consider the two-components of the modified Tsallis distribution and modified Weibull distribution for $e^{+}e^{-}$ data.~The probability functions for the two cases can be derived from the equations~(7)~and~(10).~The fit parameters, $\chi^{2}/ndf$ and $p$ values for both modified Weibull and modified Tsallis distributions are given in Tables~3,~6~and~7.~Figure~5~and~6 show the modified distributions for the L3 and OPAL data.

~One can see that values given in Table~3 show that following the two-component fits improve substantially the description.~The modified Tsallis distributions describe the data extremely well at all energies.~For the modified Weibull distribution, though the $\chi^{2}/ndf$ values improve substantially, it still fails for energies above the $Z^{0}$ peak, namely 161,~188.6,~189~and~206.2~GeV data samples for which $CL < 0.1 \%$. 

~Figure 7 shows plots of the $q$, $q_{1}$ and $q_{2}$ values estimated from the Tsallis fits and modified Tsallis fits.~The bands shown are the confidence bands.~The mean values are $q=1.388 \pm 0.095$, $q_{1}=1.077 \pm 0.017$ and $q_{2}=1.489 \pm 0.100$.~The figure shows that the $q$ values in both the cases of the Tsallis and modified Tsallis fits, exceed unity, emphasising the non-extensive nature of the Tsallis entropy.

~Figure 8 shows the dependence of $\lambda$, $\lambda_{1}$, $\lambda_{2}$, on c.m.s energy of $e^{+}e^{-}$ collisions.~As the parameter $\lambda$ is connected with the average multiplicity, it is expected to increase with the energy of collision and hence charged multiplicity.~The dependence of  $\lambda$ on c.m. energy $\sqrt{s}$ is studied as a power law $\lambda = A\sqrt{s}^{B}$, $A$ and $B$ are the fit parameters.~One finds the following values of the parameters $A$ and $B$:
 
for $\lambda$    :  $A = 10.05 \pm 0.99 \hspace*{0.5 cm} {\rm and} \hspace*{0.5 cm} B = 0.20 \pm 0.019$,

for $\lambda_{1}$ :  $A =  5.50 \pm 1.26 \hspace*{0.5 cm} {\rm and} \hspace*{0.5 cm} B = 0.29 \pm 0.04$,

and for $\lambda_{2}$ :  $A =  3.58 \pm 0.62 \hspace*{0.5 cm} {\rm and} \hspace*{0.5 cm} B  = 0.45 \pm 0.03$.

~The modified analyses for $\overline{p}p$ could not be carried out as the 2-jet fractions for these data in different rapidity intervals as well as in the full phase space are not available.

\subsection{Discussion}

~The standard Boltzmann-Gibbs statistics, with $q$=1 produces a distribution which is not consistent with the experimental data at high energy and at high multiplicity tail.~It is known that it could lead to unphysical results for systems having long-range interactions.~This is the non perturbative regime of QCD which plays an important role in hadronization.~The large particle density fluctuations [\refcite{Ada},\refcite{Bur}] present in hadron production, which could not be explained, mandated an alternative formalism to be developed. One such case is the Tsallis formalism.

In the Tsallis distribution, $K$-parameter in eq.(4) is related to $q$, the entropic index, $v_{0}$, the excluded volume associated to a particle and the volume of the system.~The detailed relationship can be found in [\refcite{TS2}].~The $K$-parameter measures the deviation from Poisson distribution caused by resonance decay and charge conservation and is related to the variance.~The definition of $K$ is motivated by the $k$ parameter of a negative binomial distribution.~The Tsallis statistics for $q > 1$ with excluded volume $v_{0}$ causes a substantial broadening of the distribution, taking it much closer to the data.~Thus the role of $K$-parameter is crucial to the value of entropic index, $q$.~In some analyses the excluded volume is fixed between 0.3-0.4 $fm^{3}$.~The corresponding value of volume $V$ then varies from few $fm^{3}$ to few tens of $fm^{3}$ [\refcite{TS2}].

The smooth increase of $q$ with the center of mass energy can be understood as the expected increase of the influence of the partonic interactions among the hadronic particles produced in the event.

~Table 2 shows that the shape parameter $k$ for the Weibull function determines the shape of the distribution.~For the highest LEP energy range, the value of $k$ does not change much within limits of errors, as can be seen in Tables~2 and ~6.~This behaviour is related to the soft gluon emission and subsequent hadronisation.~Within a given rapidity interval, the value of $k$ decreases slightly with increasing energy. However, it increases considerably from smaller to larger rapidity intervals, as can be seen in Table~4.~The scale parameter $\lambda$ of the Weibull distribution determines the width of the distribution.~Larger value of scale parameter produces a broader distribution.~This again is observed in Tables~2 and 4.~The width of the probability distribution depends upon c.m. energy.~At higher collision energies, more number of high multiplicity events are produced and the multiplicity distribution becomes broader.~As a result, $\lambda$ increases to compensate for the width.~This trend is endorsed by $\lambda$ values in the two Tables.~Similar results are observed from the modified Weibull fit distribution parameters in Table~6 where $\lambda$ values increase systematically from lower to higher rapidity windows.

~For various energies, it is shown that by appropriately weighting the multiplicity distribution with the 2-jet fraction in $e^{+}e^{-}$ collisions at $\sqrt{s}$ = 91 to 206 GeV, both the Tsallis distributions and the Weibull distributions substantially improve the agreement with the data giving the statistically significant results.~These modified Tsallis distributions reproduce the data well with $CL > 0.1 \%$ for all energies, from both the L3 and OPAL experiments.~The modified Weibull distributions, also improve the fits by several orders, but fail to describe the data at most of the energy points, as can be observed from $p$ values in Table~3

~The highest energy data sets have typically low statistics, due to which fit parameters suffer from large errors, especially for modified Tsallis distribution.~Nevertheless the present detailed analysis establishes that the performance of the Tsallis function is better than Weibull function.~The $q$ value known as entropic index for the Tsallis distribution, accounts for the non-extensive thermo-statistical effects in hadron production and is expected to exceed unity in the Tsallis statistics.~This is confirmed from the mean values of $q$, as measured for the total sample of events, events with two-jets and events with multijets.~The behaviour is found to be more pronounced in the events with higher multiplicity, as evident from the difference in q-values for 2-jet and multi-jet components.~The analyses for the Tsallis distribution and the Weibull distribution for $\overline{p}p$ collisions from 200 to 900 in full phase space as well as in restricted rapidity windows have also been done.~It is found that the Weibull distribution is excluded in full phase space at all energies and in several rapidity windows.~Whereas for the Tsallis, it is acceptable at most of the energy and rapidity values with the exception of 900 GeV in full phase space and at 540 GeV for $|y| < 0.5$ GeV where the fits have $CL < 0.1\%$.~For Tsallis distribution, in each case $q$ value exceeds unity, emphasising the non-extensive nature of the Tsallis entropy.

\section{Conclusions}
Analyses of the multiplicity distributions measured in $e^{+}e^{-}$ collisions at LEP at energies,~$\sqrt{s}$=91 to 206 GeV and in $\overline{p}p$ collisions at 200 to 900 GeV have been done in the context of Weibull and Tsallis distributions.~It is observed that the the Weibull fits for the data from L3 experiment at all energies from 130.1~GeV to 206.2~GeV for the $e^{+}e^{-}$ collisions are statistically excluded with $CL < 0.1 \%$, and for the Tsallis fits the data only at 200.2~GeV and 206.2~GeV are statistically excluded and are good for all other energies.~For the data from OPAL experiment, Weibull fits are statistically excluded for all energies between 91~GeV to 189~GeV with $CL < 0.1 \%$ in $e^{+}e^{-}$  annihilation, except the 172 GeV data.~In contrast, the Tsallis fits are excluded only for energy at 91~GeV and remain good for all other energies of 133~GeV to 189~GeV.~Tsallis non-extensive statistics gives a good description of the particle production in high energy collisions. 

{\bf Acknowledgement:} We thank Professor Edward K.G. Sarkisyan for important suggestions and comments.~One of the authors, S. Sharma is grateful to the Department of Science and Technology, Government of India for the fellowship grant.

\begin{figure}[ht]
\includegraphics[width=4.0 in, height =2.3 in]{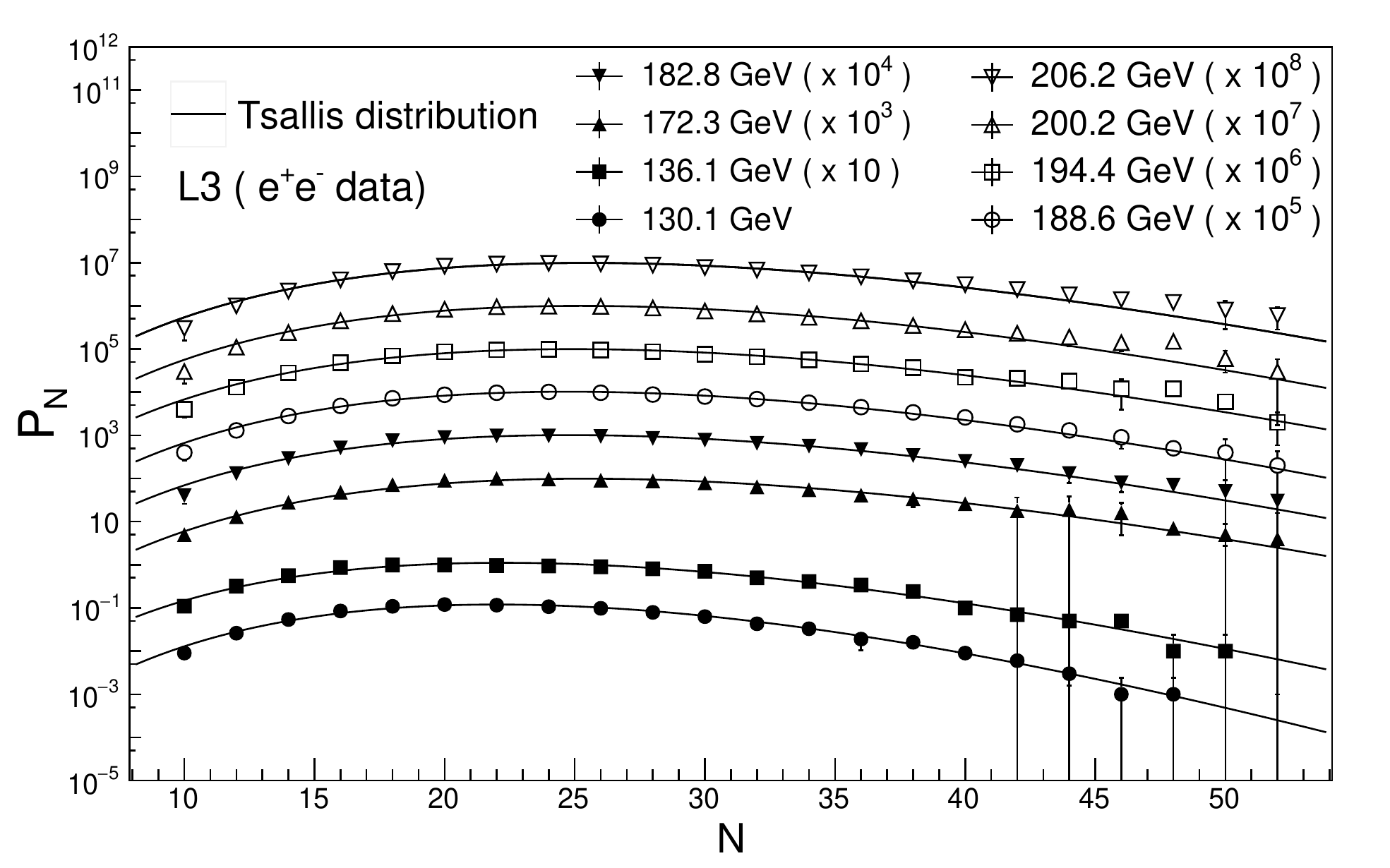}
\includegraphics[width=4.0 in, height =2.3 in]{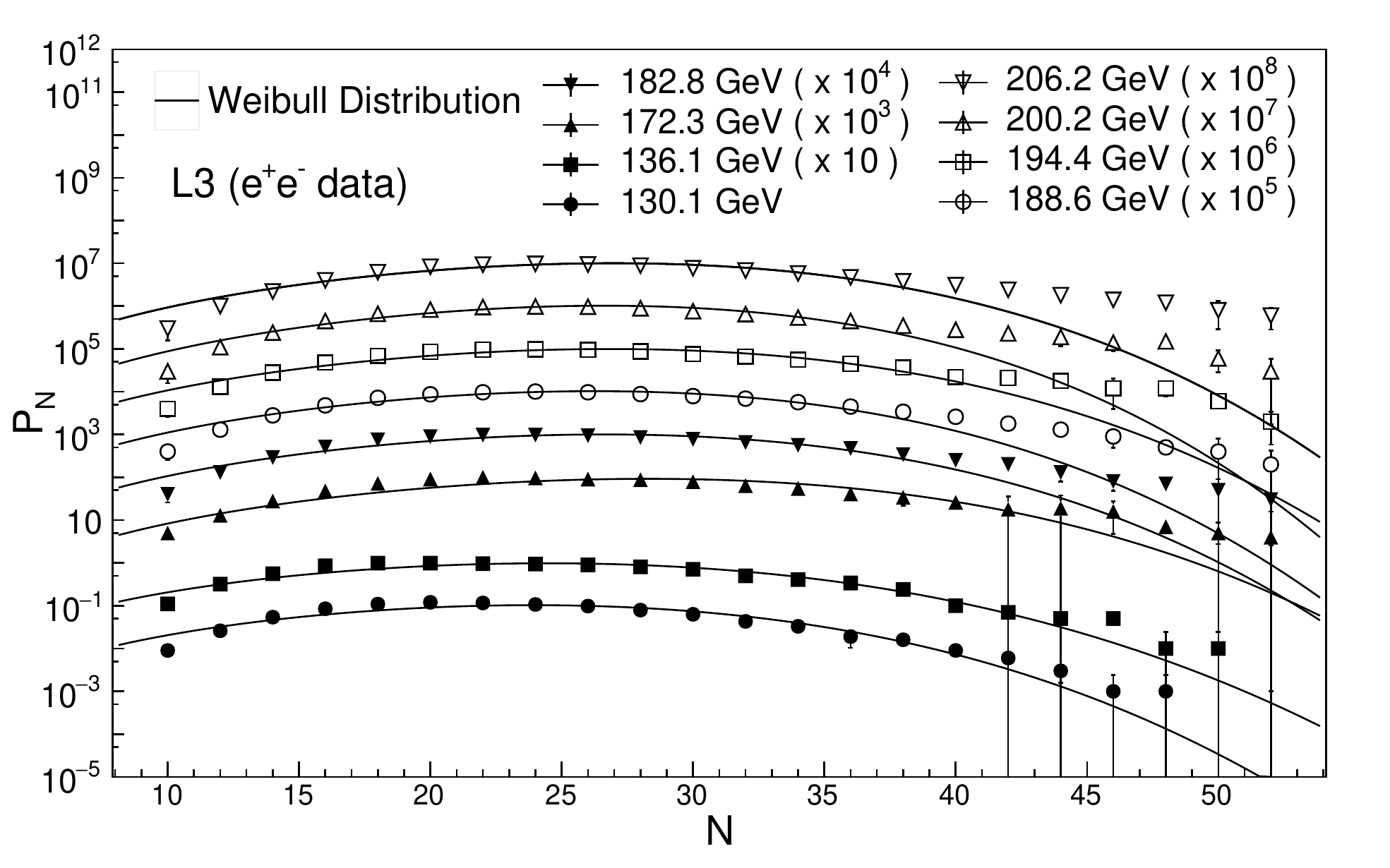}
\caption{The charged multiplicity distributions measured in $e^{+}e^{-}$ annihilation by the L3 experiment and the fits by the Tsallis distribution and Weibull distribution.}
\end{figure}
\begin{figure}[ht]
\includegraphics[width=4.0 in, height =2.3 in]{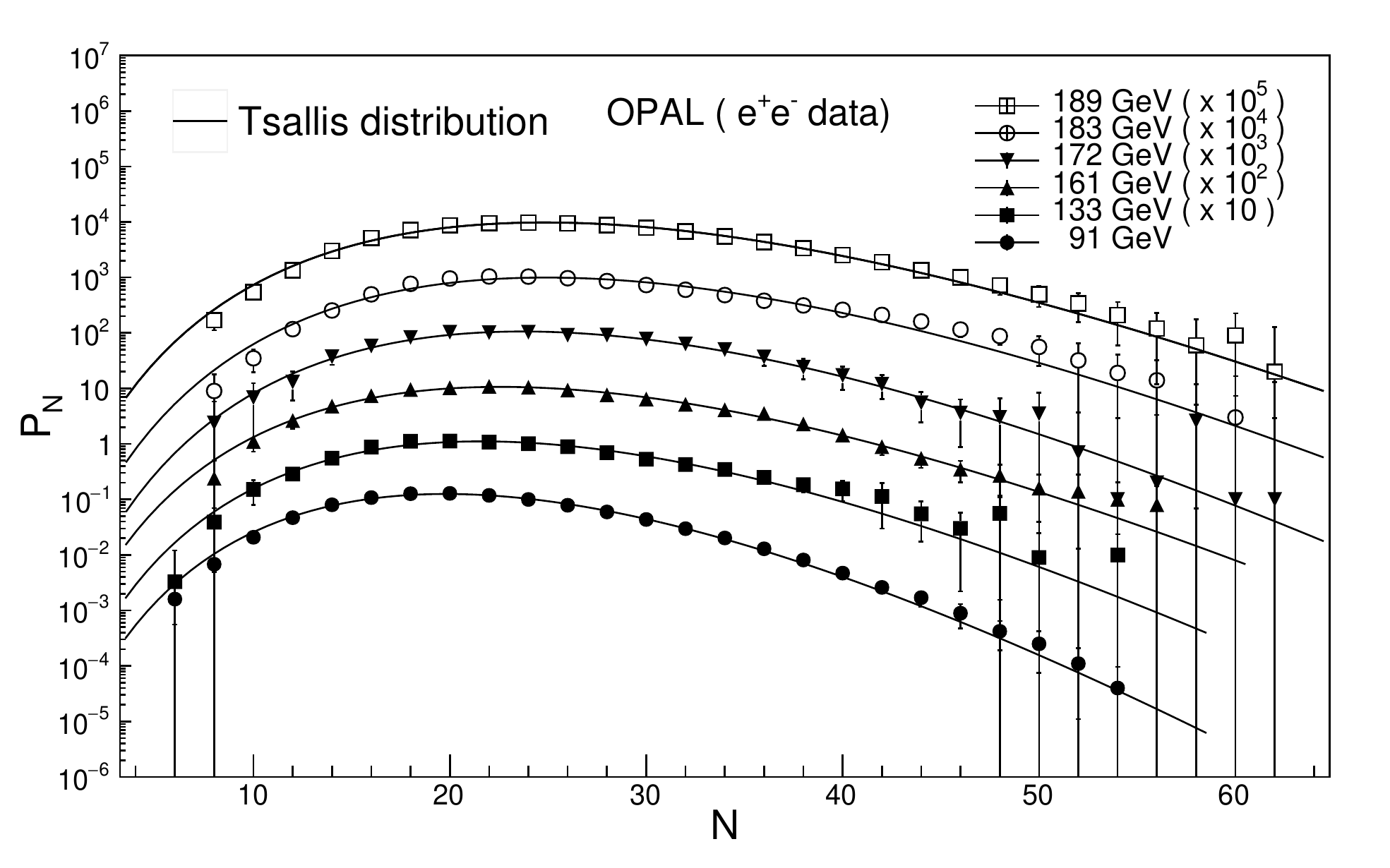}
\includegraphics[width=4.0 in, height =2.3 in]{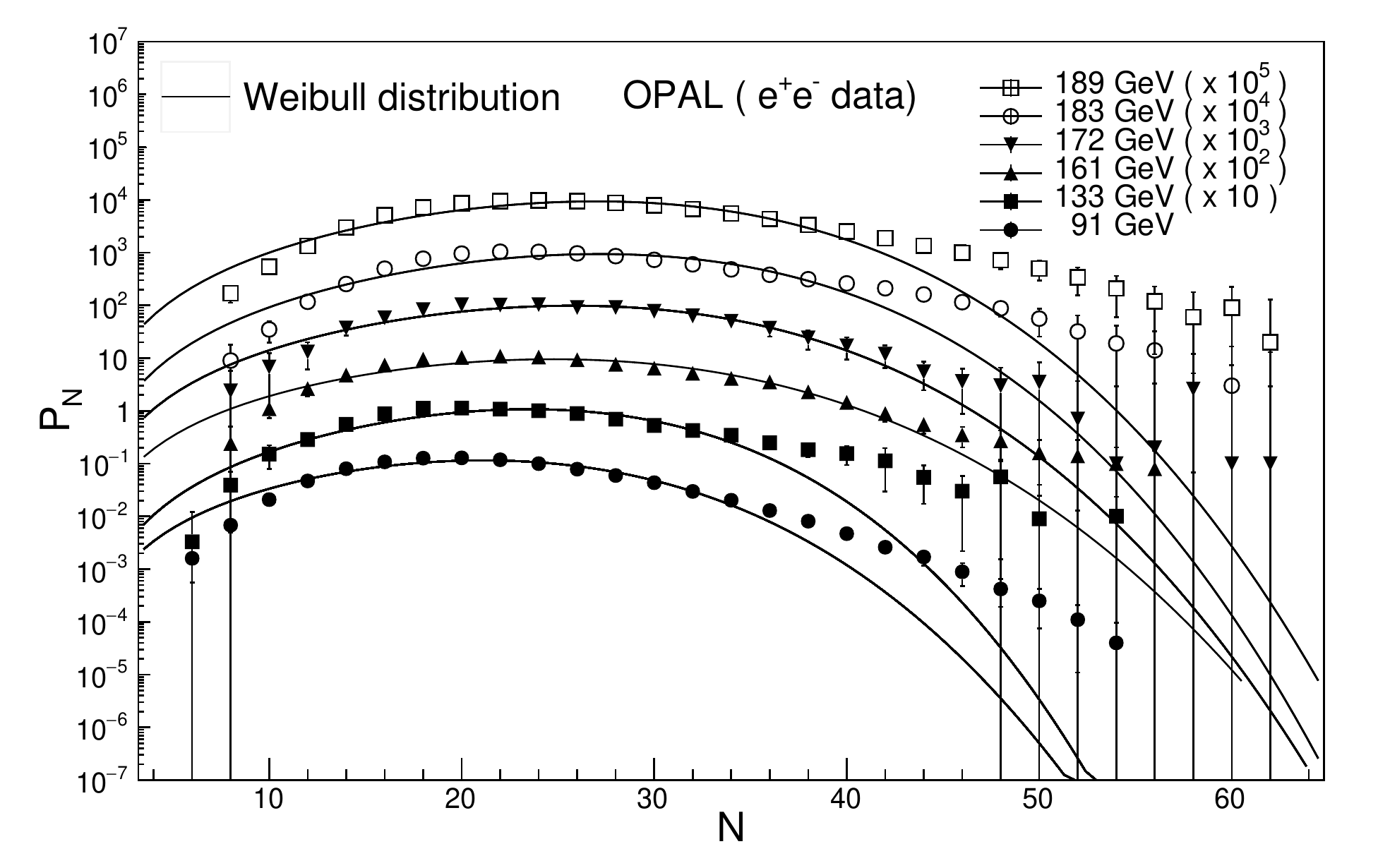}
\caption{The charged multiplicity distributions measured in $e^{+}e^{-}$ annihilation by the OPAL experiment and the fits by the Tsallis distribution and Weibull distribution.}
\end{figure}
\begin{figure}[th]
\includegraphics[width=4.0 in, height =2.3 in]{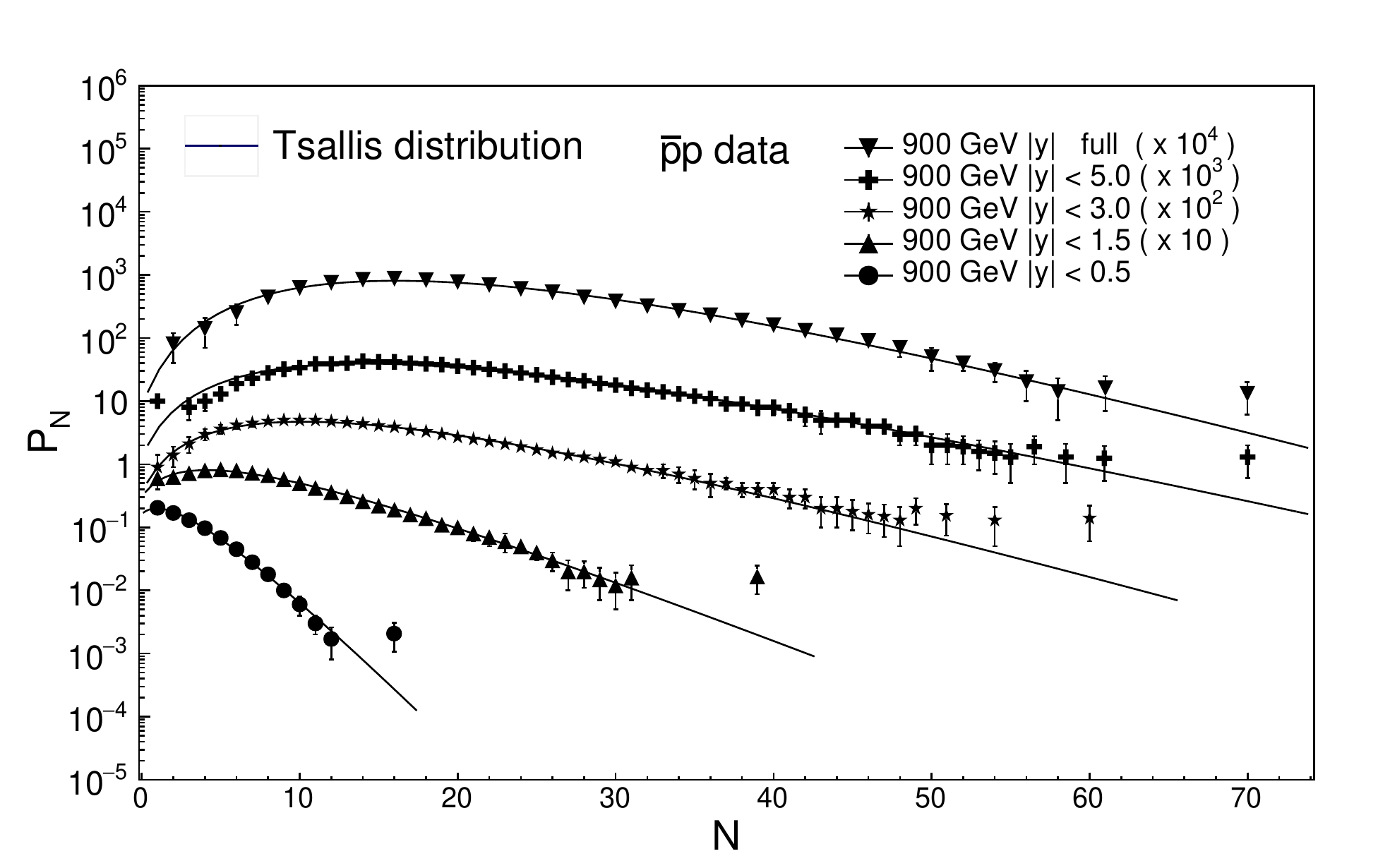}
\includegraphics[width=4.0 in, height =2.3 in]{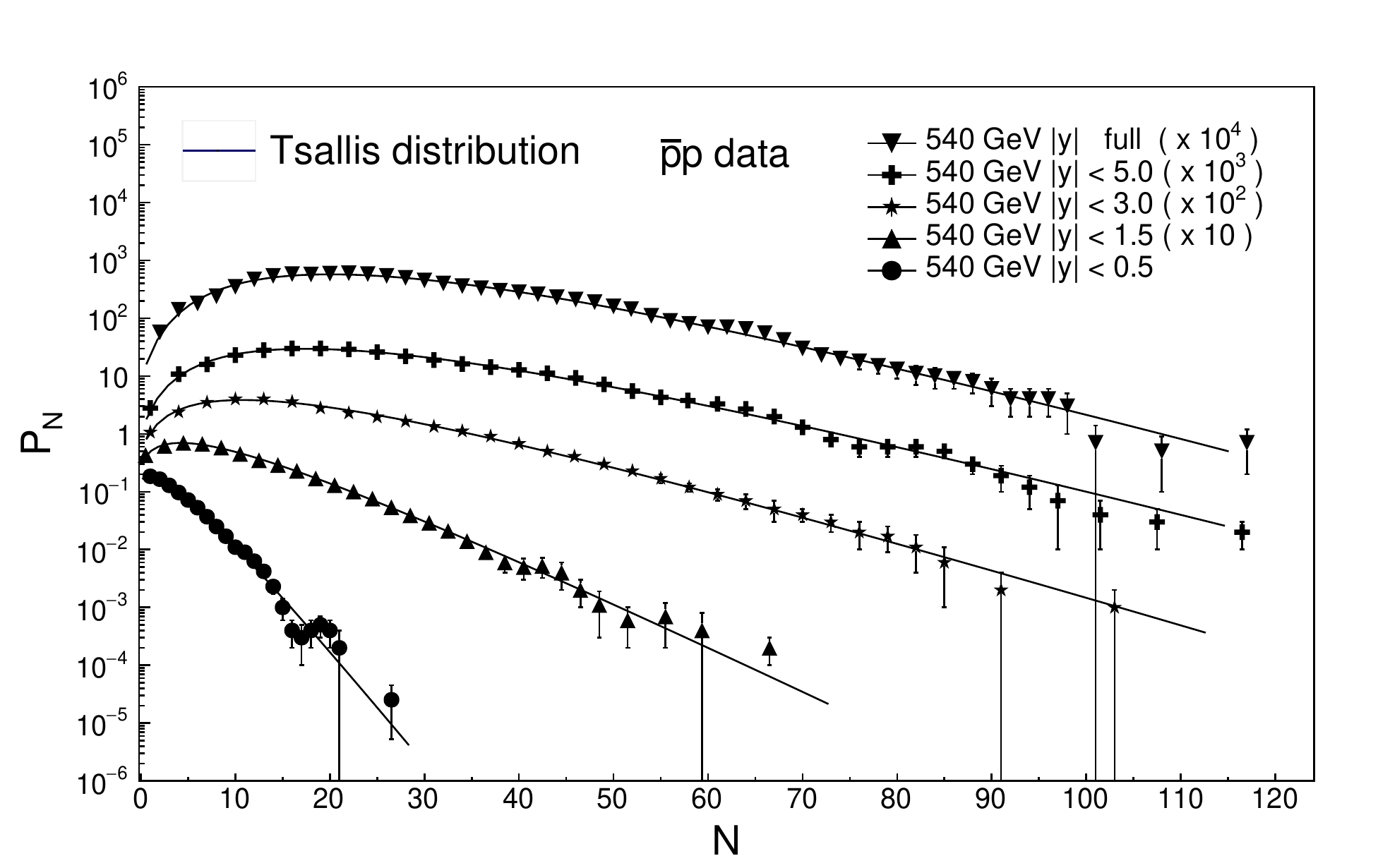}
\includegraphics[width=4.0 in, height =2.3 in]{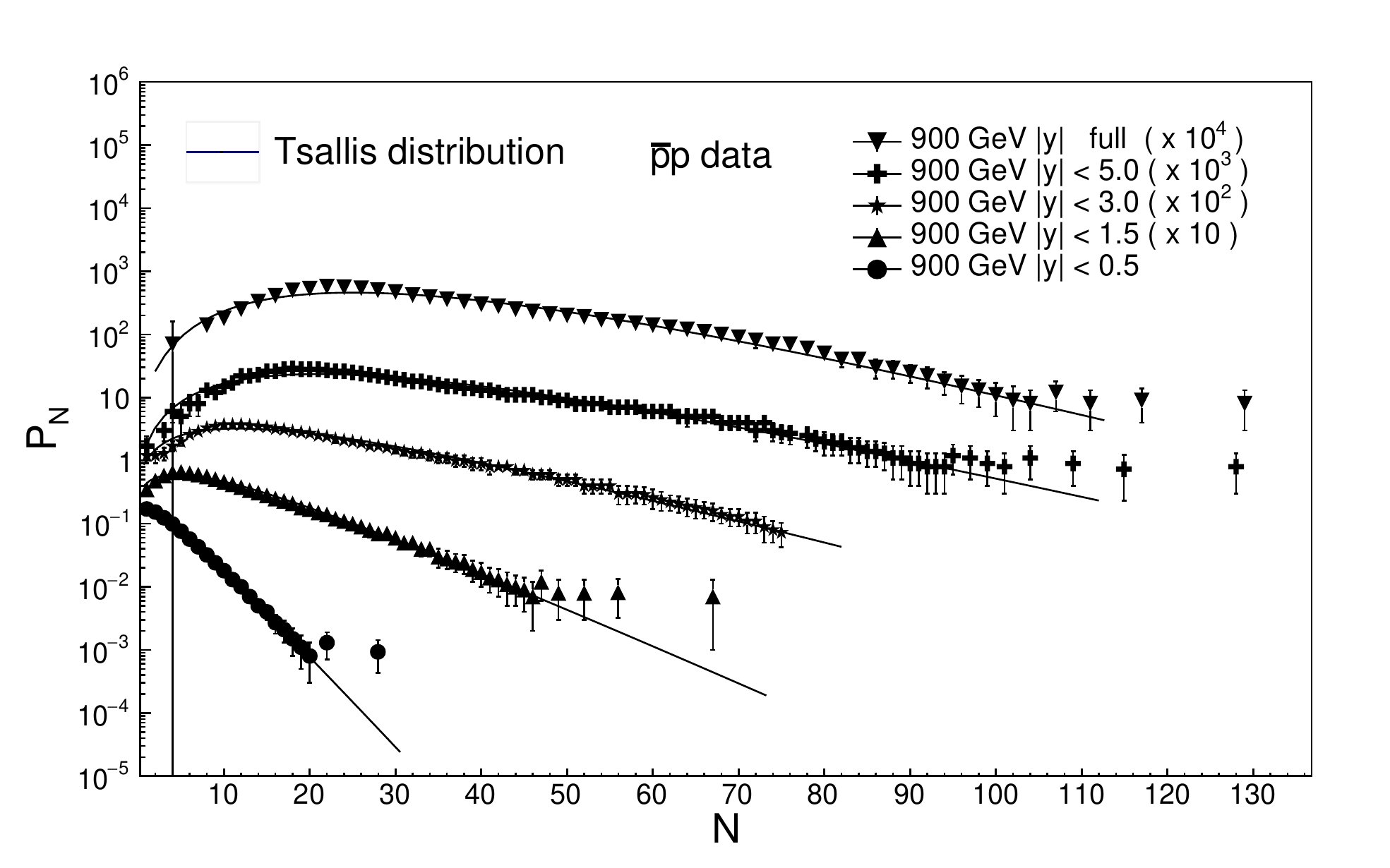}
\caption{The charged multiplicity distribution measured in $\overline{p}p$ interactions by the UA5 experiment and the fits by the Tsallis distributions.}
\end{figure}
\begin{figure}[th]
\includegraphics[width=4.0 in, height =2.3 in]{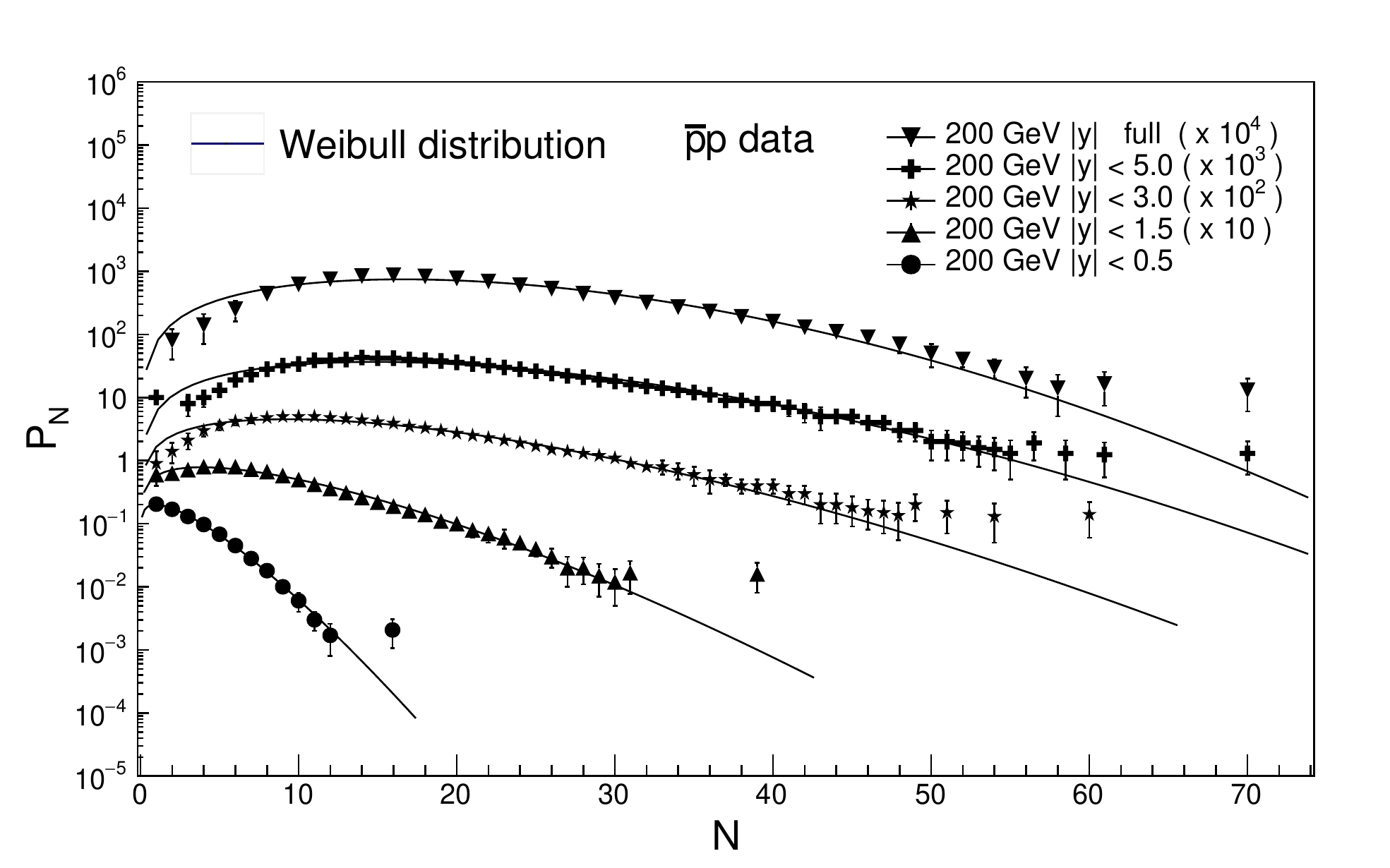}
\includegraphics[width=4.0 in, height =2.3 in]{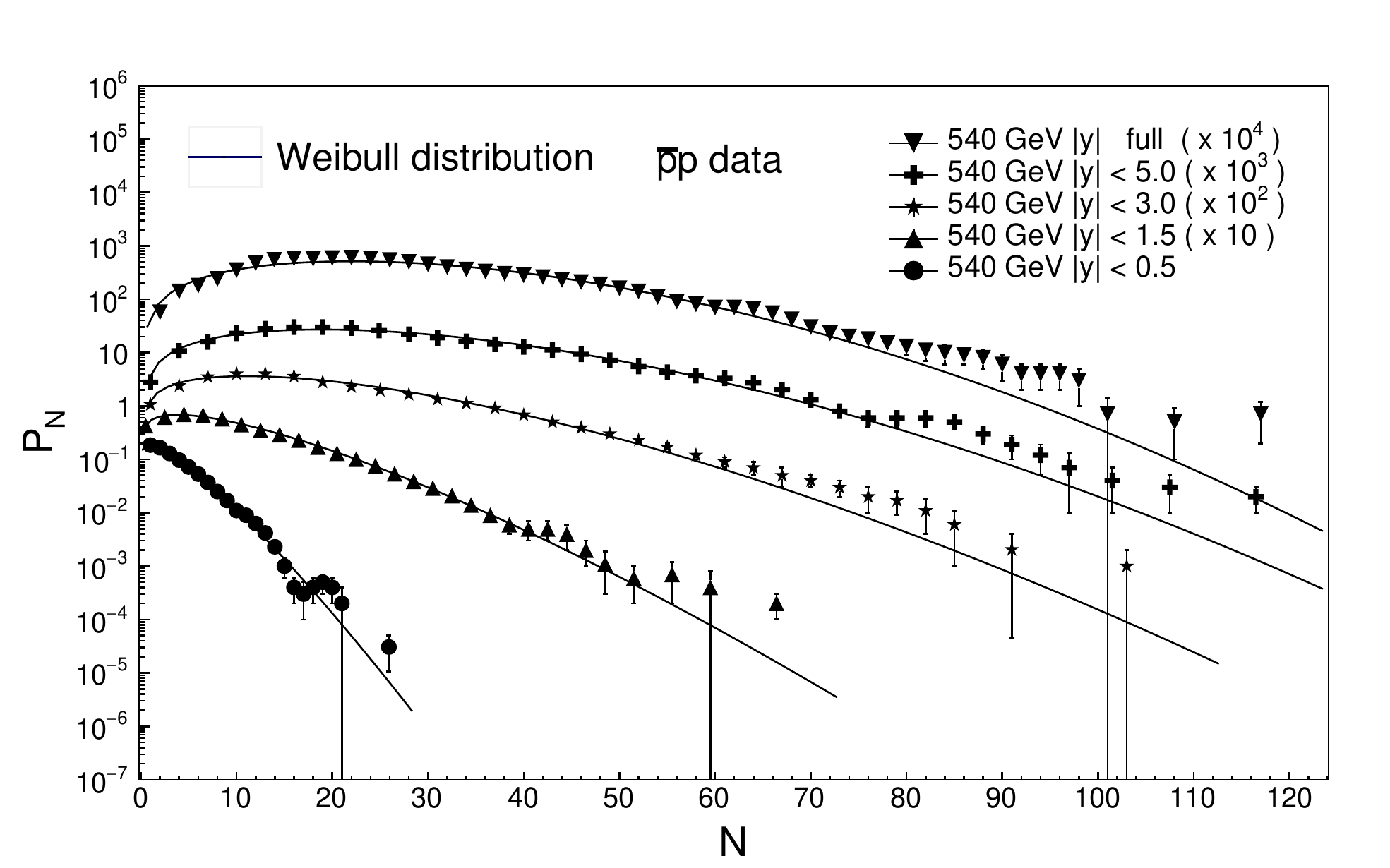}
\includegraphics[width=4.0 in, height =2.3 in]{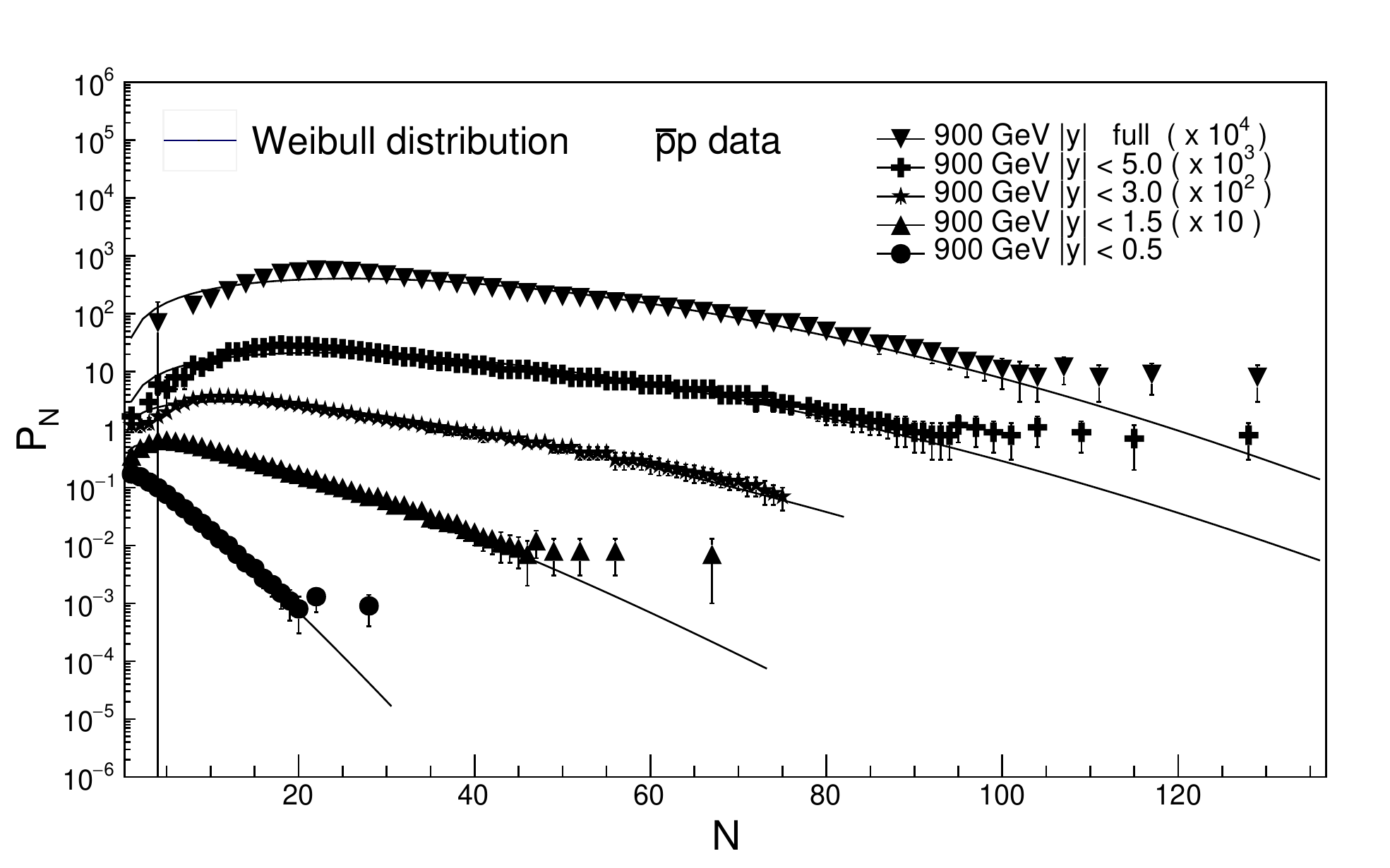}
\caption{The charged multiplicity distribution measured in $\overline{p}p$ interactions by the UA5 experiment and the fits by the Weibull distributions.}
\end{figure}

\begin{figure}[th]
\includegraphics[width=4.0 in, height =2.3 in]{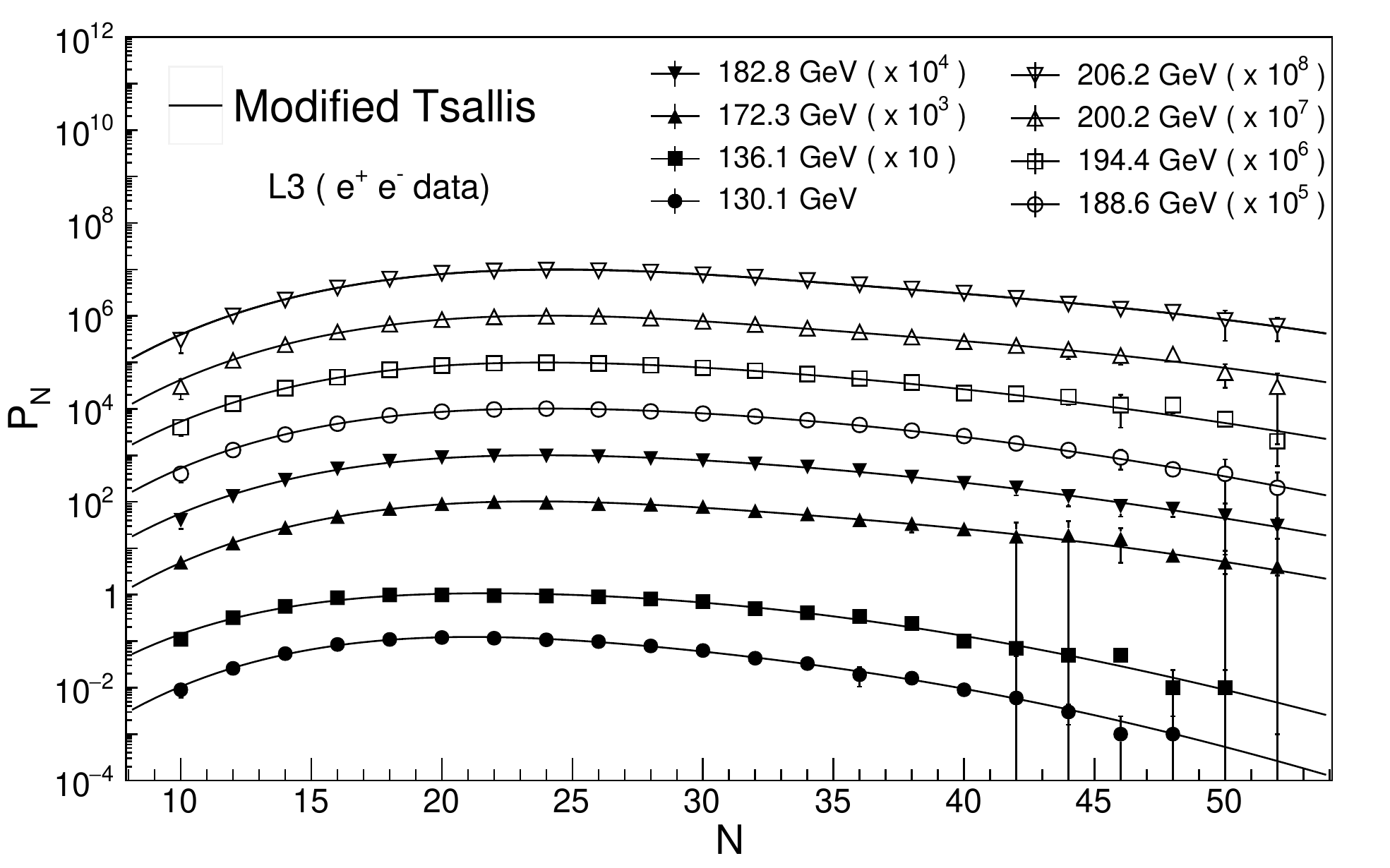}
\includegraphics[width=4.0 in, height =2.3 in]{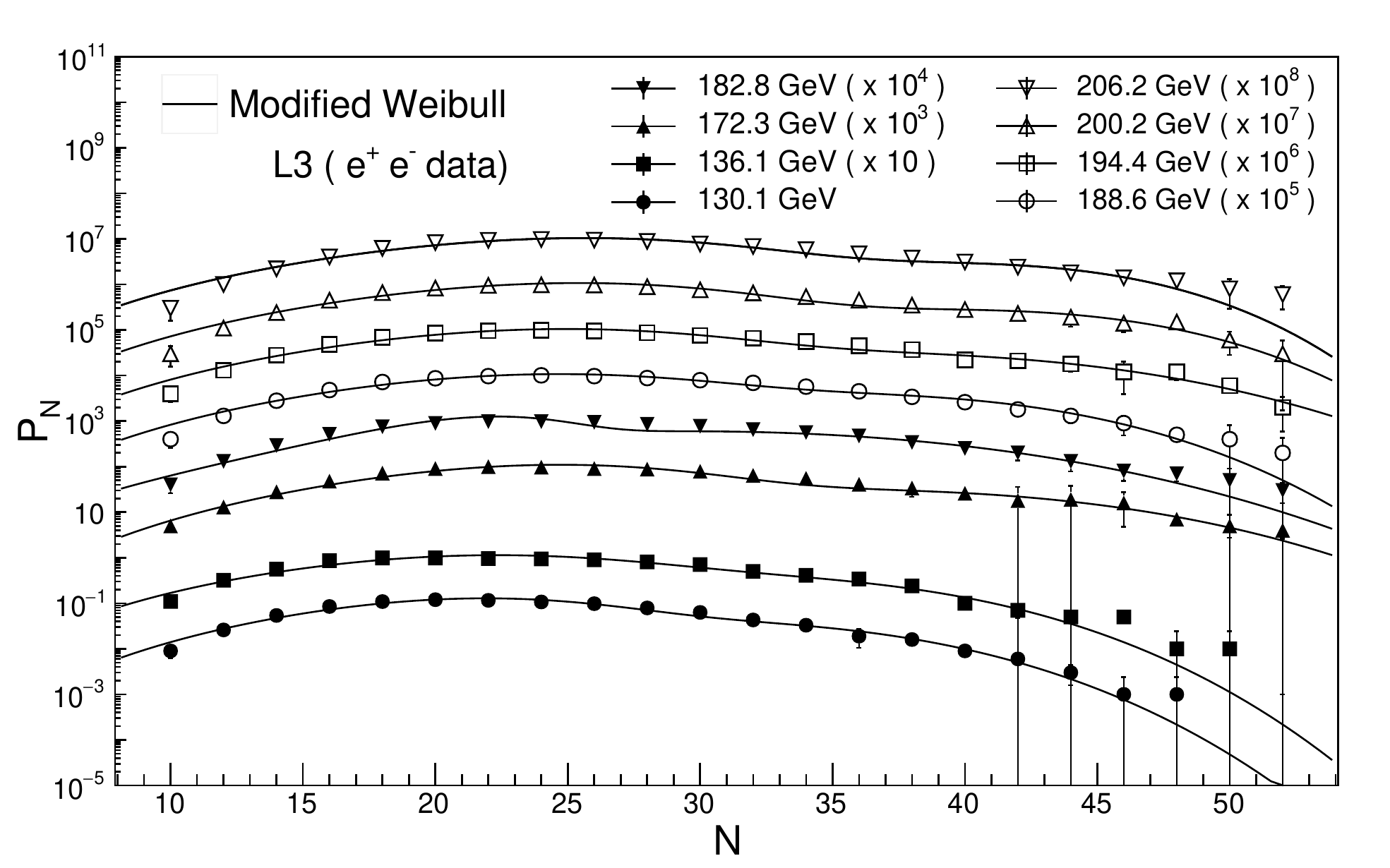}
\caption{The charged particle multiplicity distributions measured in $e^{+}e^{-}$ collisions by the L3 experiment with the fits by the modified Tsallis and Weibull distributions.}
\end{figure}
\begin{figure}[th]
\includegraphics[width=4.0 in, height =2.3 in]{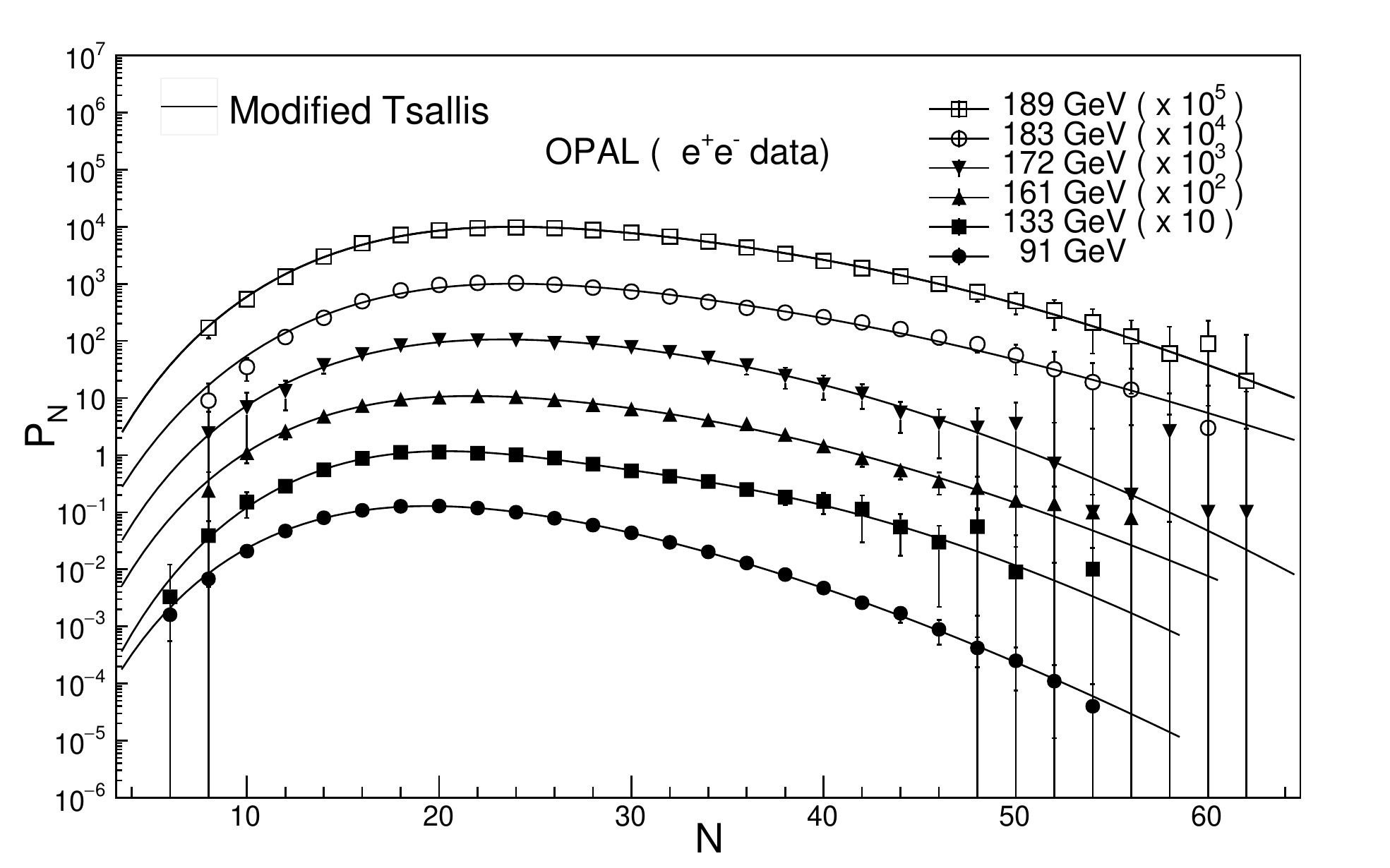}
\includegraphics[width=4.0 in, height =2.3 in]{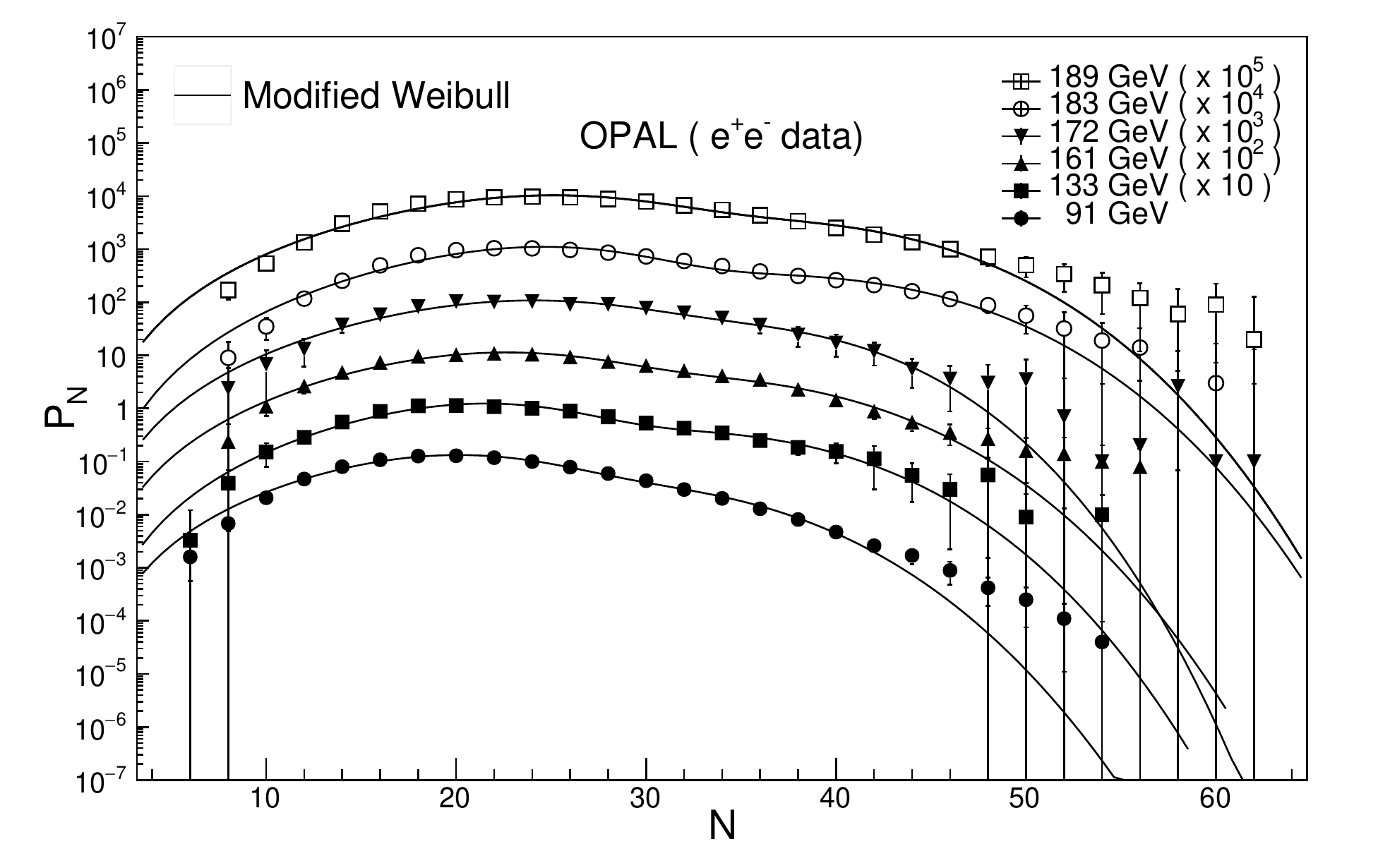}
\caption{The charged particle multiplicity distributions measured in $e^{+}e^{-}$ collisions by the OPAL experiment with the fits by the modified Tsallis and Weibull distributions.}
\end{figure}
\begin{figure}[th]
\centerline{\includegraphics[width=4.0 in, height =2.2 in]{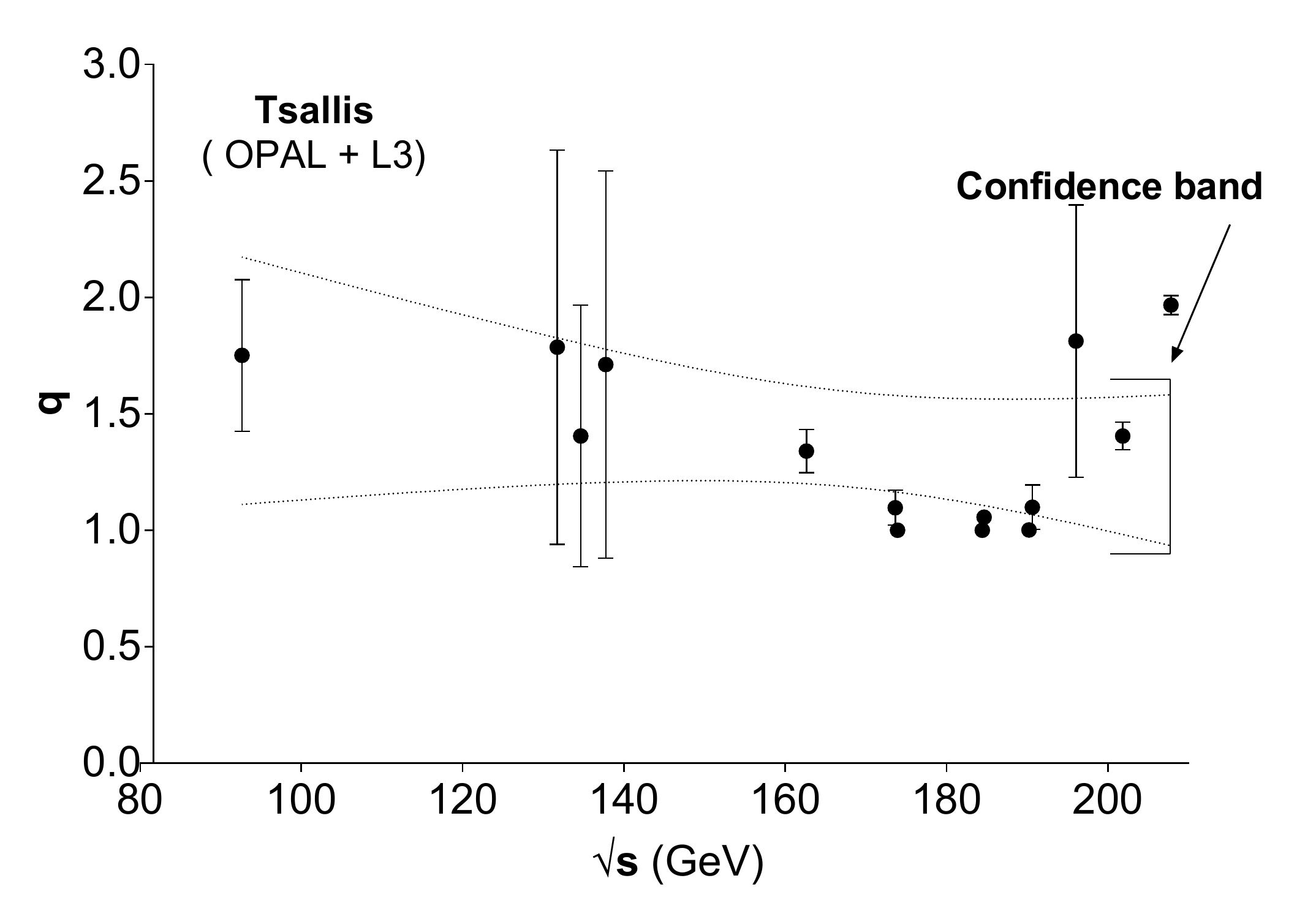}}
\centerline{\includegraphics[width=4.0 in, height =2.2 in]{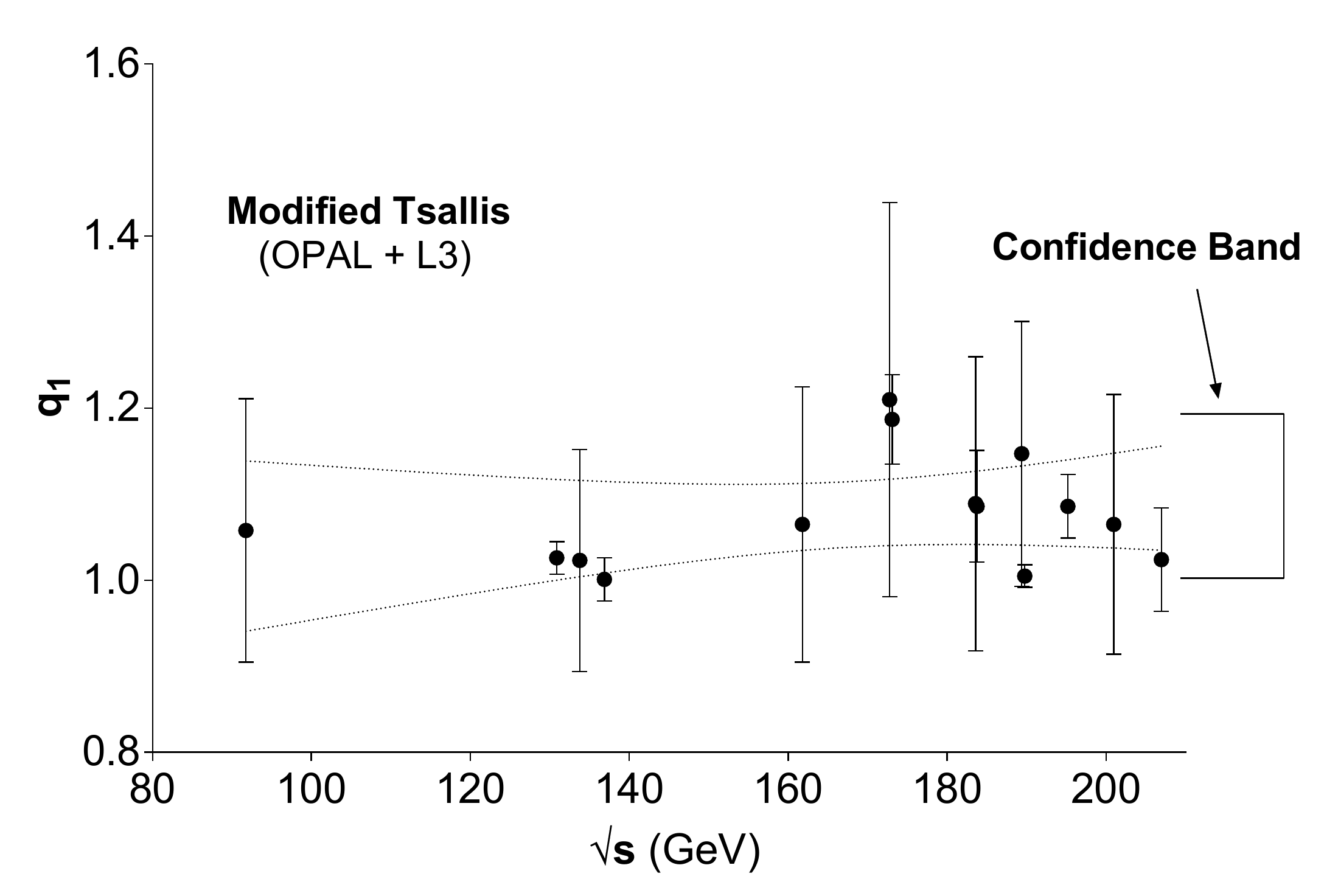}}
\centerline{\includegraphics[width=4.0 in, height =2.2 in]{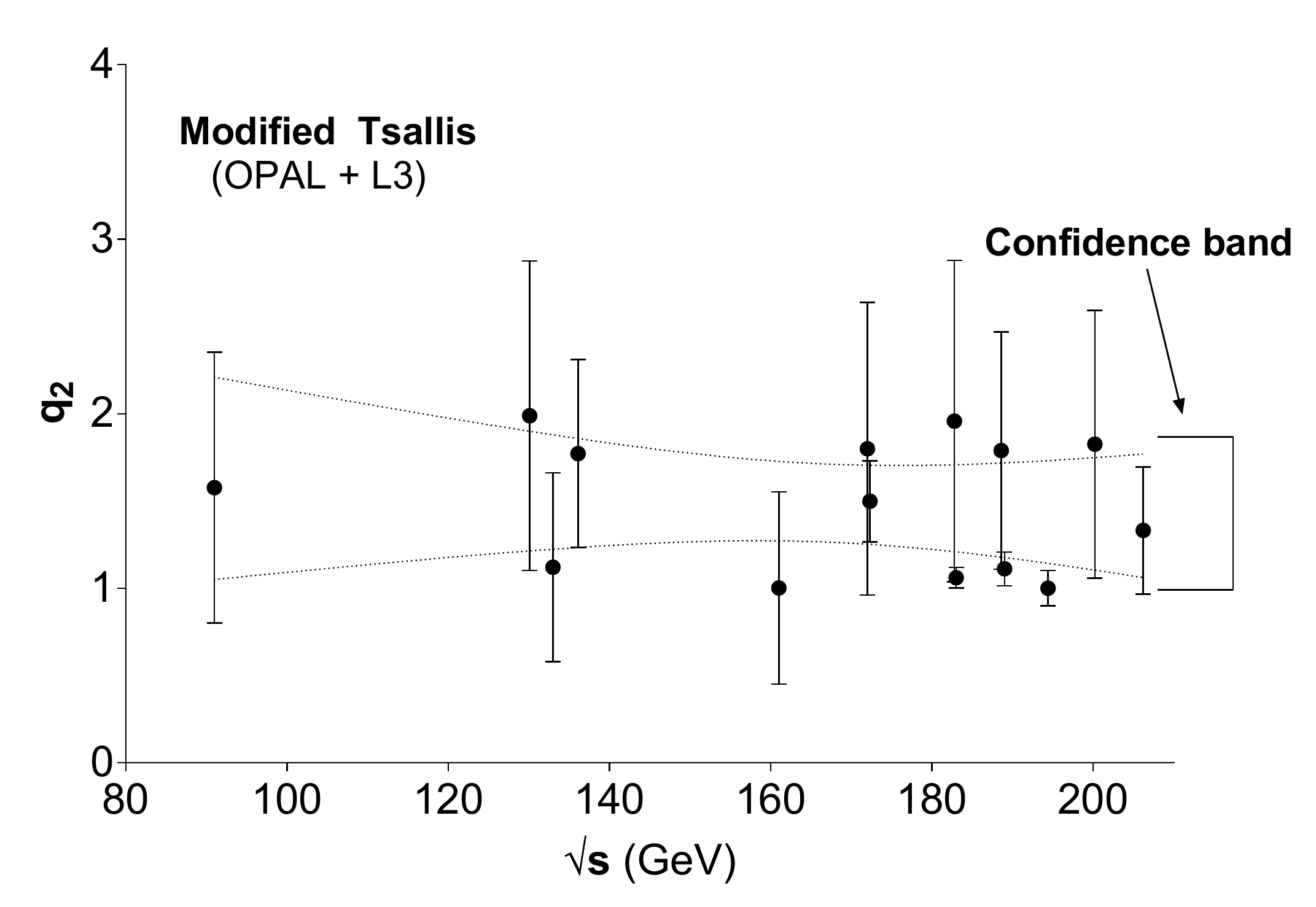}}
\caption{The Tsallis $q$, $q_{1}$ and $q_{2}$ fit parameters obtained for $e^{+}e^{-}$ data.}
\end{figure}  
\begin{figure}[th]
\centerline{\includegraphics[width=4.0 in, height =2.3in]{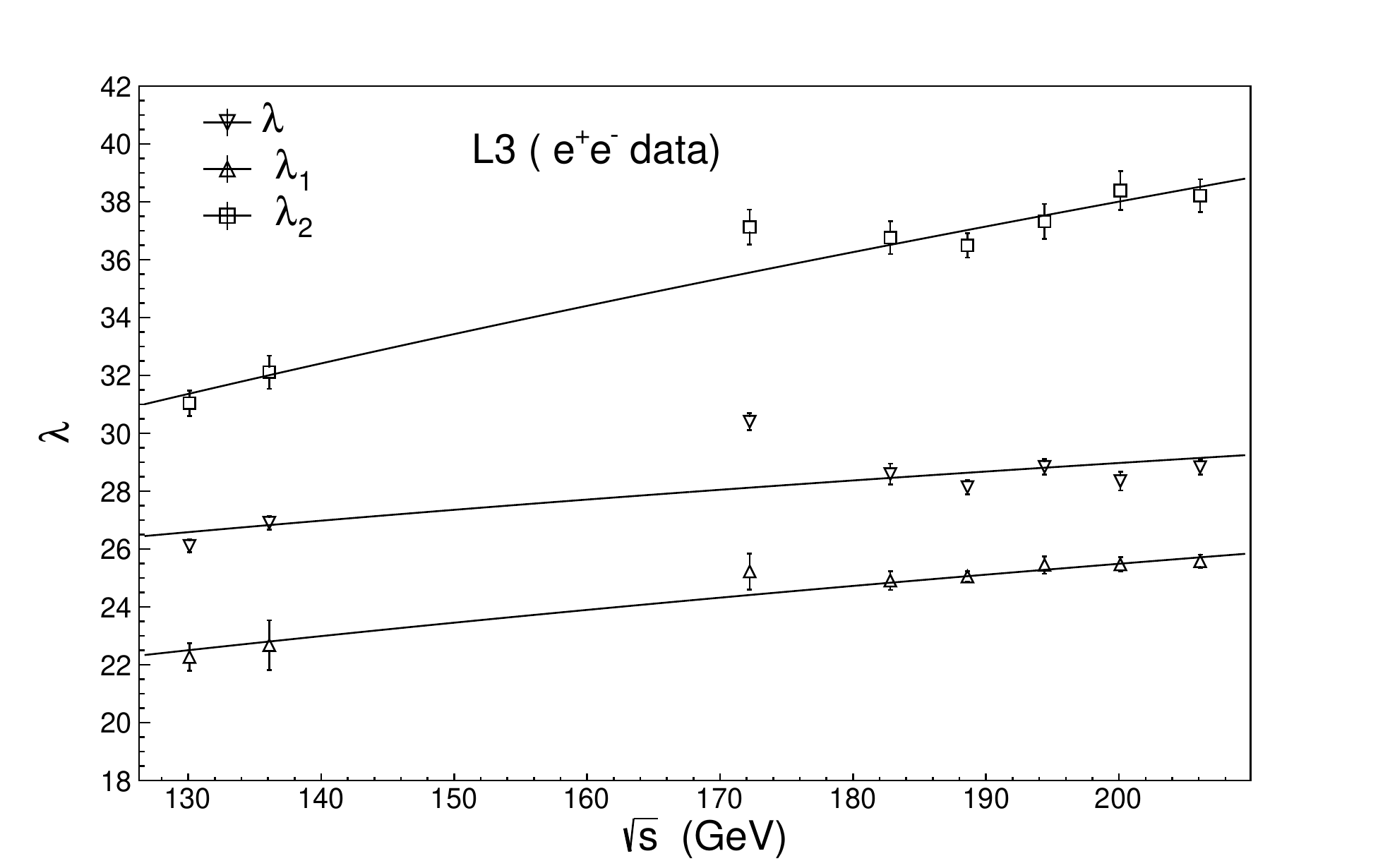}}
\caption{The Weibull distribution $\lambda$, $\lambda_{1}$ and $\lambda_{2}$ fit parameters obtained for $e^{+}e^{-}$ annihilation as a function of collision energy.~The lines represent the power law fits.}
\end{figure}

\small
\begin{table}[pt]
\tbl{Description of the data samples used in the analysis.$\alpha$-values represent the fraction of the 2-jet events in the samples as obtained by the measurements.}
{\begin{tabular}{|c|c|c|c|c|c|}
\hline
Energy   & Experiment	&[Reference] 	& No. of events  & $\alpha$(2-jet fraction) & Reference\\  
(GeV)    & 				& 	 & 				 &							&for $\alpha$\\\hline
		 				\multicolumn{6}{|c|}{$e^{+}e^{-}$}   						 \\\hline
91 & OPAL	&[\refcite{OPAL91}] & 82941    & 0.657 					& [\refcite{OPAL172}] \\\hline
133 & ''		&[\refcite{Alex}]  & 13665    & 0.662 				& [\refcite{Alex}] \\\hline
161 & ''		&[\refcite{OPAL161}]  & 1336     & 0.635 				& [\refcite{OPAL161}]  \\\hline
172 & ''		&[\refcite{OPAL172}]  & 228      			& 0.666 & [\refcite{Acc}]\\\hline
183 & ''		& "  & 1098  & 0.675 & [\refcite{Jon}]\\\hline
189 & '' 	& " & 3277     & 0.662 &  [\refcite{OPAL172}]\\\hline

 & &  &  & &  \\
130.1 & L3	&[\refcite{L3}] & 556  & 0.654 & [\refcite{L3}] \\\hline

136.1 & ''  &" & 414   & 0.649 & " \\\hline

172.3 & ''  &" & 325    &0.657 & " \\\hline
182.8 & ''  &" & 1500 & 0.668 & " \\\hline

188.6 & ''  &" & 4479   & 0.670 & "\\\hline

194.4 & ''  &"  & 2403  & 0.679 & " \\\hline
200.2 & ''  &"  & 2456    & 0.661 & "\\\hline
206.2 & ''  &"  & 4146   & 0.666 &  "  \\\hline
 
			 		\multicolumn{6}{|c|}{$\overline{p}p$ }   		 \\\hline 
        
   200 	& UA5	&[\refcite{UA51}]	 	& 4156    & -- & --\\\hline
   
   540 	& UA5	&[\refcite{UA52}] 	& 6839    & -- &	 --\\\hline
   
   900 	& UA5	&[\refcite{UA51}]	& 7344    & -- &	 --\\\hline
   
\end{tabular}}
\end{table}

\small
\begin{table}
\tbl{The parameters of the Weibull and the Tsallis functions for $e^{+}e^{-}$ collisions.}
{\begin{tabular}{|c|c|c|c|c|c|c|}
\hline

     \multirow{3}{*}{Energy}  & \multicolumn{3}{|c|}{Weibull distribution}    &   \multicolumn{3}{|c|}{Tsallis distribution}  \\
      &	 \multicolumn{3}{|c|}{}			       &   \multicolumn{3}{|c|}{}	            \\\cline{2-7}
     

 (GeV) & $k$       & $\lambda$    & $\chi^{2}/ndf$&  $K$ & $q$ & $\chi^{2}/ndf$ \\
 
  &	         &	           &			       &    &   &                 \\\hline    
 
91     &	3.548  $\pm$  0.033 & 23.197  $\pm$  0.073 & 434.91/22 & 	21.397  $\pm$  0.670	& 1.751  $\pm$  0.325 & 33.13/20\\\hline

133  & 4.029  $\pm$  0.122	& 25.239  $\pm$  0.374 & 66.88/22 &	19.891  $\pm$  2.251	& 1.405  $\pm$  0.562	& 12.67/20 \\\hline

161  & 3.542  $\pm$  0.098 &	27.174  $\pm$  0.300 & 48.11/22 &  17.983  $\pm$  1.587 & 1.340  $\pm$  0.093 & 5.81/20 \\\hline

172  & 3.813  $\pm$  0.162 & 27.942  $\pm$  0.479	& 17.87/22 &  21.274  $\pm$  3.003 &	1.097  $\pm$  0.075	& 3.94/20 \\\hline

183  & 4.069  $\pm$  0.088 & 29.117  $\pm$  0.277 & 159.61/25 & 	19.749  $\pm$  1.339 & 1.056  $\pm$  0.023 & 35.13/23 \\\hline

189  & 3.994  $\pm$  0.058 & 29.018  $\pm$  0.197 & 183.63/25 &  17.795  $\pm$  0.858 & 1.099  $\pm$  0.096 & 15.88/23 \\\hline

	 &	  &			 &					 &	      &   &\\
130.1  & 3.655  $\pm$  0.095 &	26.110  $\pm$  0.215 &	52.53/19 &	23.513  $\pm$  2.031 &	1.786  $\pm$  0.846 &	6.42/17 \\\hline

136.1  &	3.513  $\pm$  0.099 &	26.913  $\pm$  0.235 &	41.64/19 &	18.309  $\pm$  1.490 &	1.712  $\pm$  0.831 &	19.26/17 \\\hline

172.3  & 4.033  $\pm$  0.121 &	30.358  $\pm$  0.296 &	64.28/19 &   18.813  $\pm$  1.324 & 1.0001  $\pm$  0.005 &	8.78/17 \\\hline

182.8	 & 4.041  $\pm$  0.101 &	28.590  $\pm$  0.358 &	98.71/19 &	18.983  $\pm$  1.332 &	1.0001  $\pm$  0.001 &	16.98/17 \\\hline

188.6	 & 4.066  $\pm$  0.078 &	28.144  $\pm$  0.251 &	157.54/19 &	19.883  $\pm$  1.044 &	1.001  $\pm$  0.0006 &	17.28/17 \\\hline

194.4	 & 3.998  $\pm$  0.084 &	28.841  $\pm$  0.266 &	108.70/19 &	18.533  $\pm$  1.223 &	1.812  $\pm$  0.585 &	19.19/17 \\\hline

200.2	 & 4.271  $\pm$  0.099	& 28.350  $\pm$  0.322 & 127.53/19 &	20.092  $\pm$  1.497 &	1.405  $\pm$  0.060 &	27.14/17 \\\hline

206.2	 & 4.151  $\pm$  0.079 &	28.831  $\pm$  0.262 &	168.91/19 &	19.631  $\pm$  1.195 &	1.967  $\pm$  0.041 &	32.41/17 \\\hline
\end{tabular}}
\end{table}

\newpage

\small
\begin{table}
\tbl{$\chi^{2}/ndf$ comparison and $p$ values for different energies and for different distributions of $e^{+}e^{-}$ annihilation.}
{\begin{tabular}{|c| c|c |c|c| c|c| c|c| c|c|}
\hline
\multirow{3}{*}{Energy}  & \multicolumn{2}{|c|}{Weibull } & \multicolumn{2}{|c|}{Tsallis }   & \multicolumn{2}{|c|}{Mod. Weibull	}  & \multicolumn{2}{|c|}{Mod. Tsallis	 }    \\
 & \multicolumn{2}{|c|}{ Distribution} & \multicolumn{2}{|c|}{ Distribution}   & \multicolumn{2}{|c|}{Distribution	}  & \multicolumn{2}{|c|}{Distribution	 }    \\\cline{2-9}

(GeV) & $\chi^2/ndf$ & p value & $\chi^2/ndf$  & p value & $\chi^2/ndf$  & p value & $\chi^2/ndf$  & p value  \\
&       & 		 & 		& 		& 		& 		& 		& 		  	 \\\hline

91 &	434.91/22 &	0.0001	 & 33.13/20	 & 0.0329	 & 12.20/20	 & 0.9090	 & 5.32/16	 & 0.9940\\\hline
133 &	66.88/22  &	0.0001	 & 12.67/20	 & 0.8899	 & 14.71/20	 & 0.7933	& 2.40/16	 & 1.0000\\\hline
161 &	48.11/22  &	0.0011	 & 5.81/20	 & 0.9991	 & 72.78/20	 & 0.0001 &	2.89/16	 & 0.9999\\\hline 
172 &	17.87/22  &	0.8466	 & 3.94/20	 & 1.0000 	& 7.83/23	 & 0.9987	& 3.31/19	 & 1.0000\\\hline
183 &	159.61/25 &	0.0001	 & 35.13/23	 & 0.0508	 & 28.80/23	 & 0.1871	& 12.91/19	 & 0.8437\\\hline
189 &	183.63/25 &	0.0001	 & 15.88/23	 & 0.8595	 & 52.11/23	 & 0.0005 &	2.73/19	 & 1.0000\\\hline  
&       & 		 & 		& 		& 		& 		& 		& 		  	 \\
130.1	& 52.53/19	& 0.0001 	& 6.42/17	& 0.9899	& 9.75/17 	& 0.9137	& 4.20/13	& 0.9889\\\hline
136.1	& 41.64/19	& 0.0002		& 19.26/17	& 0.3138	& 27.63/17	& 0.0495	& 16.61/13	& 0.2178\\\hline
172.3	& 64.28/19	& 0.0001	& 8.78/17	& 0.9471		& 11.26/17	& 0.8427	& 1.51/13	& 1.0000\\\hline
182.8	& 98.71/19	& 0.0001	& 16.98/17	& 0.4557	& 29.92/17	& 0.0269	& 5.72/13	& 0.9558\\\hline
188.6	& 157.54/19	& 0.0001	& 17.28/17	& 0.4356	& 40.93/17	& 0.0010		& 5.83/13	& 0.9521\\\hline
194.4	& 108.70/19	& 0.0001	& 19.19/17	& 0.3177	& 30.86/17	& 0.0208	& 6.94/13	& 0.9052\\\hline
200.2	& 127.53/19	& 0.0001	& 27.14/17	& 0.0562		& 29.01/17	& 0.0344	& 4.61/13	& 0.9828\\\hline
206.2	& 168.91/19	& 0.0001	& 32.41/17	& 0.0134	& 41.41/17	& 0.0008	& 4.12/13	& 0.9898\\\hline
\end{tabular}}
\end{table}

\small
\begin{table}
\tbl{The Parameters of the Weibull and the Tsallis functions for $\overline{p}p$ collisions.}
{\begin{tabular}{|c|c|c|c|c|c|c|c|}
\hline
\multirow{2}{*}{Energy}  &\multirow{2}{*}{$|y|$}  &\multicolumn{3}{|c|}{Weibull Distribution } &  \multicolumn{3}{|c|}{Tsallis Distribution } \\\cline{3-8}
   &	&  &			 &					 &	  &      &\\     
   	(GeV) & & k  & $\lambda$ & $\chi^{2}/ndf$  &  K & q & $\chi^{2}/ndf$ \\
  &	&  &			 &					 &	  &      &\\\hline
  
    \multirow{5}{*}{200}  & 0.5	&1.278 $\pm$ 0.039	&3.171 $\pm$ 0.076	& 5.37/10& 2.278 $\pm$ 0.257	&1.310 $\pm$ 0.006 &6.35/8\\\cline{2-8}
       &1.5	&1.420 $\pm$ 0.029	&9.078 $\pm$ 0.147 	&16.61/29 & 2.314 $\pm$ 0.120	&1.138 $\pm$ 0.025 & 11.41/27\\\cline{2-8}
 
 & 3.0	&1.618 $\pm$ 0.033	&16.740 $\pm$ 0.208 	&41.89/49 &3.201 $\pm$ 0.038	&1.021 $\pm$ 0.015& 19.78/47\\\cline{2-8}
 
 & 5.0	&1.843 $\pm$ 0.034	&22.660 $\pm$ 0.271 	&69.38/55 &3.497 $\pm$ 0.160	&1.008 $\pm$ 0.002&62.51/53\\\cline{2-8}

& full	&2.001 $\pm$ 0.043	&23.410 $\pm$ 0.285  	&44.71/28 &4.518 $\pm$ 0.230	&1.002 $\pm$ 0.001&10.72/26\\\hline
 &	          &			 &	     &   &   & & \\
   	 	 
 \multirow{5}{*}{540} & 0.5	&1.218 $\pm$ 0.018	&3.587 $\pm$ 0.048  	& 19.84/19&1.697 $\pm$ 0.045	&1.428 $\pm$ 0.021 &29.15/18\\\cline{2-8}
 
 & 1.5	&1.371 $\pm$ 0.002	&10.530 $\pm$ 0.046     	&24.77/26 &1.995 $\pm$ 0.044	&1.184 $\pm$ 0.019 &13.41/24\\\cline{2-8}
 
 & 3.0	&1.572 $\pm$ 0.013	&20.920 $\pm$ 0.161    	&119.50/28 &2.482 $\pm$ 0.043	&1.057 $\pm$ 0.005 &33.86/26\\\cline{2-8}
 
 & 5.0	&1.804 $\pm$ 0.015	&29.580 $\pm$ 0.205   	&127.13/33 & 3.115 $\pm$ 0.052	&1.013 $\pm$ 0.004 &39.51/31\\\cline{2-8}

& full	&1.938 $\pm$ 0.017	&31.940 $\pm$ 0.206 	&164.99/49 & 3.623 $\pm$ 0.063	&1.009 $\pm$ 0.003 &46.41/47
\\\hline

 &	          &			 &	  &	     &   &&\\
\multirow{5}{*}{900} & 0.5	&1.140 $\pm$ 0.026	&4.075 $\pm$ 0.084    	&6.49/19 & 1.474 $\pm$ 0.100	&1.504 $\pm$ 0.007&5.58/17\\\cline{2-8}
 & 1.5	&1.292 $\pm$ 0.022	&12.260 $\pm$ 0.183    	&37.05/48 & 1.787 $\pm$ 0.067	&1.221 $\pm$ 0.079&22.88/46\\\cline{2-8}
 
 & 3.0	&1.454 $\pm$ 0.022	&24.460 $\pm$ 0.307    	&77.52/72 & 2.169 $\pm$ 0.071	&1.098 $\pm$ 0.017 &41.06/70\\\cline{2-8}
 
 & 5.0	&1.749 $\pm$ 0.025	&36.370 $\pm$ 0.424    	&141.81/97 &2.886 $\pm$ 0.086	&1.024 $\pm$ 0.002&63.09/93\\\cline{2-8}

& full	&1.849 $\pm$ 0.030	&39.420 $\pm$ 0.415    	&162.01/51 &3.562 $\pm$ 0.112	&1.011 $\pm$ 0.003&73.29/47\\\hline  
 
\end{tabular}}
\end{table}

\small
\begin{table}
\tbl{$\chi^{2}/ndf$ comparison and $p$ values for different energies and for different rapidities for $\overline{p}p$ collisions.}
{\begin{tabular}{|c|c| c|c |c|c|}
\hline
\multirow{3}{*}{Energy} & \multirow{3}{*}{$|y|$}  & \multicolumn{2}{|c|}{Weibull Distribution }   & \multicolumn{2}{|c|}{Tsallis Distribution }  \\
&  & \multicolumn{2}{|c|} 	{}			& \multicolumn{2}{|c|}	{} \\\cline{3-6} 

(GeV)& & $\chi^2/ndf$ & p value & $\chi^2/ndf$  & p value  \\\hline

\multirow{5}{*}{200}   &0.5    &5.37/10	&0.8651	&6.35/8	  &0.6047\\\cline{2-6}
		&1.5	&16.61/29	&0.9679	&11.41/27	&0.9963\\\cline{2-6}
		&3.0	&41.89/49	&0.7543	&19.78/47	&0.9998\\\cline{2-6}
		&5.0	&69.38/55	&0.0919	&62.51/53	&0.1745\\\cline{2-6}
		&full	&44.71/28	&0.0236	&10.72/26 &0.9964\\\hline

 &         &  			  & & &          \\					
					
\multirow{5}{*}{540}    &0.5 &19.84/19	 &0.4043&29.15/18&0.0466\\\cline{2-6}
		&1.5	&24.77/26	    &0.532	&13.41/24	&0.9589\\\cline{2-6}
		&3.0	&119.50/28	&0.0001	&33.86/26	&0.1386\\\cline{2-6}
		&5.0	&127.13/33	&0.0001	&39.51/31	&0.1405\\\cline{2-6}
		&full	&164.99/49	&0.0001	&46.41/47	&0.4969\\\hline

 &         &  			  & & &          \\
 					
\multirow{5}{*}{900} &0.5   &6.49/19 &0.9965	&5.58/17 &0.9938\\\cline{2-6}
		&1.5	&37.05/48	    &0.8742	&22.88/46	&0.9985\\\cline{2-6}
		&3.0	&77.52/72	    &0.3071	&41.06/70	&0.9977\\\cline{2-6}
		&5.0	&141.81/97	&0.0021	&63.09/93	&0.9909\\\cline{2-6}
		&full	&162.01/51	&0.0001	&73.29/47	&0.0084\\\hline

\end{tabular}}
\end{table}
\small
\begin{table}[pt]
\tbl{The Parameters of the modified Weibull function for $e^{+}e^{-}$ collisions.}
{\begin{tabular}{|c|c|c|c|c|c|}
\hline
Energy   & $k_{1}$ & $\lambda_{1}$ & $k_{2}$ & $\lambda_{2}$  & $\chi^{2}/ndf$   \\
  (GeV)       & 	  	   &               &	        &	             &	                \\\hline    
91  & 4.556  $\pm$  0.228 & 23.030  $\pm$  0.471 & 4.684  $\pm$  0.482 & 32.810  $\pm$  0.544	& 12.20/20	\\\hline

133 &	4.810  $\pm$  0.192 &	22.050  $\pm$  0.418 &	4.811  $\pm$  0.557 &	32.480  $\pm$  0.775	& 14.71/20  \\\hline

161  &	4.412  $\pm$  0.063 &	20.620  $\pm$  0.119 &	4.221  $\pm$  0.133 &	28.401  $\pm$  0.168	& 72.78/20  \\\hline

172 	& 4.537  $\pm$  0.307 & 24.380  $\pm$  0.691 & 5.363  $\pm$  0.963 &	34.070  $\pm$  0.928	& 7.83/23 \\\hline

183 	& 5.025  $\pm$  0.139 & 25.170  $\pm$  0.310 & 5.157  $\pm$  0.358 & 37.510  $\pm$  0.516	& 28.80/23\\\hline

189 	& 4.629  $\pm$  0.094 &	25.570  $\pm$  0.260 & 5.121  $\pm$  0.377 &	36.690  $\pm$  0.405	& 52.11/23\\\hline

    & 	  			 	 &	                     &	                    &	                     &   \\ 
						
130.1 & 4.811  $\pm$  0.248 & 22.267  $\pm$  0.469 &	4.696  $\pm$  0.352 & 31.041  $\pm$  0.438 & 9.75/17 \\\hline

136.1	 & 4.286  $\pm$  0.318	& 22.681  $\pm$  0.858 &	4.808  $\pm$  0.698 &	32.110  $\pm$  0.575 &	27.63/17\\\hline

172.3	 & 5.054  $\pm$  0.218 &	25.191  $\pm$  0.624 &	4.631  $\pm$  0.318 &	37.133  $\pm$  0.602 &	11.26/17	 \\\hline

182.8 	& 4.757  $\pm$  0.140	& 24.910  $\pm$  0.322	& 5.315  $\pm$  0.497 &	36.765  $\pm$  0.570	& 29.92/17	 \\\hline

188.6 	& 4.721  $\pm$  0.096	& 25.051  $\pm$  0.184 & 5.445  $\pm$  0.326 & 36.492  $\pm$  0.422	& 40.93/17  \\\hline

194.4	 & 4.695  $\pm$  0.131 &	25.454  $\pm$  0.299 &	4.659  $\pm$  0.300 &	37.322  $\pm$  0.601	& 30.86/17	 \\\hline

200.2 	& 4.915  $\pm$  0.120 &	25.473  $\pm$  0.248	& 5.197  $\pm$  0.398 & 38.391  $\pm$  0.670	& 29.01/17\\\hline

206.2 	 & 4.846  $\pm$  0.106 &	25.582  $\pm$  0.226 & 5.602  $\pm$  0.437 &	38.210  $\pm$  0.564	& 41.41/17\\\hline 
\end{tabular}}
\end{table}

\small
\begin{table}[pt]
\tbl{The Parameters of the modified Tsallis function for $e^{+}e^{-}$ collisions.}
{\begin{tabular}{|c|c|c|c|c|c|c|c|}
\hline
Energy   & $K_{1}$ & $q_{1}$ &  $K_{2}$ &  $q_{2}$ &  $\chi^{2}/ndf$ \\
 (GeV)	& &	  		 &		&		&	 	     \\\hline    
91  &	 22.140  $\pm$  2.022   & 1.058  $\pm$  0.153	 & 50.001  $\pm$  3.202 	& 1.577  $\pm$  0.776	& 5.32/16\\\hline 
133 &	 68.540  $\pm$  31.590  & 1.023  $\pm$  0.129	& 46.340  $\pm$  37.620 	& 1.120  $\pm$  0.541	& 2.40/16\\\hline 
161 &	 24.890  $\pm$  5.287	& 1.065  $\pm$  0.160   & 50.001  $\pm$  35.600 	& 1.002  $\pm$  0.551	& 2.89/16\\\hline 
172 &	 33.920  $\pm$  11.590	& 1.210  $\pm$  0.229   & 50.002  $\pm$  36.680 	& 1.799  $\pm$  0.838	& 3.31/19\\\hline 
183 &	 17.970  $\pm$  2.197	& 1.086  $\pm$  0.065   & 50.001  $\pm$   3.275 	        & 1.061  $\pm$  0.059	& 12.91/19\\\hline 
189 &	 32.101  $\pm$  5.396	& 1.005  $\pm$  0.013   & 37.410  $\pm$  12.660  	& 1.112  $\pm$  0.097	& 2.73/19\\\hline 

& 	  &				&		&	 &	    \\						
130.1  	& 25.032  $\pm$  4.022	& 1.026  $\pm$  0.019   & 50.027  $\pm$  28.561 	& 1.988  $\pm$  0.887	& 4.20/13\\\hline 
136.1  	& 24.591  $\pm$  3.979	& 1.001  $\pm$  0.025	& 50.021  $\pm$  27.911 &  1.772  $\pm$  0.539	& 16.61/13\\\hline 
172.3 	& 33.573  $\pm$  5.151	& 1.187  $\pm$  0.052	& 50.007  $\pm$  38.722 &  1.498  $\pm$  0.232	& 1.51/13\\\hline 
182.8  	& 31.388  $\pm$  3.787	& 1.089  $\pm$  0.171	& 50.001  $\pm$  27.303 &  1.958  $\pm$  0.921	& 5.72/13\\\hline 
188.6   & 30.031  $\pm$  2.654	& 1.147  $\pm$  0.154	& 50.004  $\pm$  8.850 &  1.789  $\pm$  0.680	& 5.83/13\\\hline 
194.4   & 31.455  $\pm$  3.500	& 1.086  $\pm$  0.037	& 50.006  $\pm$  37.741 & 1.001  $\pm$  0.100	& 6.94/13\\\hline 
200.2  	& 34.391  $\pm$  3.687	& 1.065  $\pm$  0.151	& 50.005  $\pm$  30.089 &  1.825  $\pm$  0.767	& 4.61/13\\\hline 
206.2  	& 33.101  $\pm$  2.983	& 1.024  $\pm$  0.060	& 50.012  $\pm$  28.161 &  1.331  $\pm$  0.364	& 4.12/13\\\hline 
\end{tabular}}
\end{table}

\end{document}